\documentclass[journal]{IEEEtran}
\IEEEoverridecommandlockouts

\usepackage{amsmath,graphicx}
\usepackage[fontsize=9pt]{scrextend}

\usepackage{booktabs}
\usepackage{balance}
\usepackage{cite}

\usepackage{amsmath,empheq, epsfig,amssymb,amsthm,url,amsfonts}
\usepackage{algorithm}
\usepackage{algpseudocode}
\usepackage{multirow}
\usepackage{mdframed}
\usepackage{url}
\usepackage{enumitem}
\usepackage[makeroom]{cancel}
\usepackage{dblfloatfix}
\usepackage{array}
\usepackage{comment}
\usepackage{epstopdf}
\usepackage{float}
\usepackage{subfig}
\usepackage{breqn}
\usepackage{xcolor}
\usepackage{cases}
\usepackage{caption}
\usepackage{tabularx}
\usepackage[printonlyused,withpage]{acronym}
\usepackage{balance}
\usepackage{pifont}
\usepackage{graphbox} 

\usepackage{hyperref}
\usepackage{bbm}
\usepackage[table]{xcolor} 
\usepackage{colortbl}
\definecolor{gray}{rgb}{0.9,0.9,0.9} 
\usepackage{tikz}
\usetikzlibrary{fit,calc}

\colorlet{mypink}{red!40}
\colorlet{myblue}{cyan!60}

\usepackage{setspace}
\setlist[itemize]{label=$\triangleright$}
\newtheoremstyle{break}
{}
{}
{\itshape}
{}
{\bfseries}
{.}
{\newline}
{}
\theoremstyle{break}

\theoremstyle{definition}

\usepackage{caption}



\newcommand{\vect}[1]{\mathbf{#1}}
\newcommand{\bs}[1]{\boldsymbol{#1}}
\newcommand{\E}{\mathbb{E}}
\usepackage{url}

\makeatletter
\def\thmhead@plain#1#2#3{%
	\thmname{#1}\thmnumber{\@ifnotempty{#1}{ }\@upn{#2}}%
	\thmnote{ {\the\thm@notefont#3}}}
\let\thmhead\thmhead@plain
\makeatother

\graphicspath{{./figs/}}

\newcommand{\argmin}{\operatornamewithlimits{argmin}}

\newcommand{\diag}{\mbox{{diag}}}

\newcommand{\Ib}{{\bf I}}

\newsavebox\mybox

\acrodef{SE}{speech enhancement}
\acrodef{AVSE}{audio-visual speech enhancement}
\acrodef{STFT}{short-time Fourier transform}
\acrodef{ESTOI}{extended short-time objective intelligibility}
\acrodef{NMF}{non-negative matrix factorization}
\acrodef{DNN}{deep neural network}
\acrodef{VAE}{Variational Auto-Encoder}
\acrodef{DKF}{deep Kalman filter}
\acrodefplural{VAEs}{variational auto-encoders}
\acrodef{EM}{expectation-maximization}
\acrodef{TF}{time-frequency}
\acrodef{ELBO}{evidence lower bound}
\acrodef{LR}{Living Room}
\acrodef{SDR}{signal-to-distortion ratio}
\acrodef{PESQ}{perceptual evaluation of speech quality}
\acrodef{SNR}{signal-to-noise ratio}
\acrodef{DNNs}{deep neural networks}
\acrodef{VESDE}{variance-exploding stochastic differential equation}
\acrodef{SDE}{stochastic differential equation}
\acrodef{GAN}{generative adversarial networks}
\acrodefplural{GANs}{generative adversarial networks}
\acrodef{SI-SDR}{scale-invariant signal-to-distortion ratio}
\acrodef{MOS}{mean opinion score}
\acrodef{SGMSE+}{score-based generative model for speech enhancement}
\acrodef{NCSNPP++}{Noise-Conditional Score Network}
\acrodef{WSJ}{Wall Street Journal}
\acrodef{UDiffSE}{Unsupervised Diffusion-Based Speech Enhancement}
\acrodef{DIFFUSER}{Diffusion-based Fast Framework for Unsupervised Speech Enhancement}
\acrodef{PC}{Predictor-Corrector}
\acrodef{DMPS}{Diffusion Model Posterior Sampling}
\acrodef{NN}{neural network}

\newcommand{\normalc}{\mathcal{N}_{\mathbb{C}}(\vect{0}, \vect{I})}

\usepackage{bm}

\usepackage{cite}
\usepackage{textcomp}
\def\BibTeX{{\rm B\kern-.05em{\sc i\kern-.025em b}\kern-.08em
    T\kern-.1667em\lower.7ex\hbox{E}\kern-.125emX}}
    
\newcommand{\cmark}{\textcolor{green!60!black}{\ding{51}}}%
\newcommand{\xmark}{\textcolor{red!70!black}{\ding{55}}}%
    
\newcommand{\cmarkgreen}{\textcolor{green!60!black}{\ding{51}}} 
\newcommand{\xmarkred}{\textcolor{red!70!black}{\ding{55}}}     

\begin{document}

\title{Diffusion-based Frameworks for Unsupervised Speech Enhancement\\

\author{Jean-Eudes Ayilo, Mostafa Sadeghi, Romain Serizel, and Xavier Alameda-Pineda, \IEEEmembership{Senior Member, IEEE}}

\thanks{J.-E. Ayilo, M. Sadeghi, and R. Serizel are with the Multispeech team, Université de Lorraine, CNRS, Inria, Loria, Nancy, France.}
\thanks{X. Alameda-Pineda is with the RobotLearn team, Université Grenoble Alpes, Inria, Grenoble, France.}

\thanks{This work was supported by the French National Research Agency (ANR) under the project REAVISE (ANR-22-CE23-0026-01).}
}

\maketitle
\begin{abstract}
This paper addresses \emph{unsupervised} diffusion-based single-channel speech enhancement (SE). Prior work in this direction combines a score-based diffusion model trained on clean speech with a Gaussian noise model whose covariance is structured by non-negative matrix factorization (NMF). This combination is used within an iterative expectation–maximization (EM) scheme, in which a diffusion-based posterior-sampling E-step estimates the clean speech. We first revisit this framework and propose to explicitly model both speech and acoustic noise as latent variables, jointly sampling them in the E-step instead of sampling speech alone as in previous approaches. We then introduce a new {semi-supervised} SE framework that replaces the NMF noise prior with a diffusion-based noise model, learned jointly with the speech prior in a single conditional score model. Within this framework, we derive two variants: one that implicitly accounts for noise and one that explicitly treats noise as a latent variable. Experiments on WSJ0--QUT and VoiceBank--DEMAND show that explicit noise modeling systematically improves SE performance for both NMF-based and diffusion-based noise priors. Under matched conditions, the diffusion-based noise model attains the best overall quality and intelligibility among unsupervised methods, while under mismatched conditions the proposed NMF-based explicit-noise framework is more robust and suffers less degradation than several supervised baselines. {Code, demo, and supplementary materials are publicly available.}\footnote{\url{https://github.com/jeaneudesAyilo/enudiffuse}}
\end{abstract}



\section{{Introduction}}\label{sec:intro}
Speech enhancement (SE) is a widely studied speech restoration task that aims to recover an underlying clean speech signal from a noisy recording. With the rise of deep neural networks (DNNs) and the development of diverse architectures and learning paradigms, substantial progress has been achieved in SE~\cite{wang2018supervised}, in both supervised and unsupervised settings. Supervised methods require clean–noisy speech training pairs, whereas unsupervised methods rely on clean speech only, noisy speech only, or unpaired clean and noisy speech data, but never on clean–noisy speech pairs. {These} broad families of SE methods can be further refined by distinguishing between \textit{generative} and \textit{non-generative} training, leading to four categories: \textit{supervised non-generative} \cite{luo2019conv,wang2023tf,10745728,wang2018supervised}, \textit{supervised generative} \cite{pascual2017segan,lu2022conditional,lemercier2023storm,zhang2025composite,shetu2025gan}, \textit{unsupervised non-generative} \cite{alamdari2021improving,10096481,wisdom2020unsupervised,tzinis2022remixit,wu2023self}, and \textit{unsupervised generative} \cite{xiang2020parallel,fu2022metricgan,bando2018statistical,bie2022unsupervised,nortier2023unsupervised,ayilo2024diffavse,sadeghi2025posterior}. The main motivation for unsupervised methods is to improve the generalization of trained DNNs to unseen (mismatched) conditions, without requiring large and diverse paired datasets, which are often impractical to collect in real-world scenarios. Nevertheless, under matched conditions, they typically exhibit a performance gap compared to their supervised counterparts ~\cite{bie2022unsupervised}.

Generative SE methods explicitly or implicitly model the prior distribution of speech and/or noise, and have recently gained renewed interest with the advent of diffusion models \cite{lemercier2025diffusion}. Indeed, the high quality of images, videos, and audio generated by diffusion models demonstrates their ability to act as powerful data priors. This has encouraged research on diffusion-based SE in both supervised and unsupervised settings. The particular case of unsupervised diffusion-based SE, which is the focus of this paper, offers the possibility of using diffusion models as strong priors over audio data (for both speech and noise), without relying on paired datasets.

An early line of work on this topic is that of Berné et al.~\cite{nortier2023unsupervised}, followed by~\cite{ayilo2024diffavse, sadeghi2025posterior}, where a score-based diffusion model trained on a clean-speech corpus is used as a generative prior for speech, while the acoustic noise prior is modeled as a Gaussian distribution with a \ac{NMF}-structured covariance.  To perform SE, an iterative Expectation–Maximization (EM) procedure is implemented, where the E-step applies diffusion posterior sampling to estimate the clean speech, and the M-step updates the noise parameters. Such a methodology provides a task-agnostic pre-trained prior, which could be reused for different restoration tasks without retraining \cite{lemercier2025diffusion}. This approach is not limited to speech processing. In image restoration, for example, a diffusion model is trained on a clean image dataset and leveraged at inference together with optimized parametric operators modeling the degradation affecting the images \cite{daras2024survey}. For blind image deblurring, Chung et al.~\cite{chung2023parallel} train two diffusion models, one serving as a clean-image prior and the other as a deblurring-kernel prior.

In the audio processing domain, similar ideas of using parallel diffusion models over both speech and acoustic noise have been proposed \cite{11464305, mariani2024multisource}. Yemini et al.~\cite{11464305} perform audio-visual speech separation in the presence of noise by training a diffusion model as a clean speech prior conditioned on lips video, and another diffusion model as an acoustic noise prior. In the context of music source separation, Mariani et al.~\cite{mariani2024multisource} train separate diffusion models for each independent source, and accordingly perform diffusion posterior sampling at inference to estimate the sources. 

In the unsupervised diffusion-based SE works \cite{nortier2023unsupervised, ayilo2024diffavse, sadeghi2025posterior}, the sampled clean speech and the estimated noise parameters during inference must jointly reconstruct the observed noisy speech to ensure data consistency in the likelihood. Noticeably, while the clean speech is sampled from its posterior distribution during the E-step, the noise is not explicitly sampled from a posterior distribution. Instead, only the NMF parameters defining the noise covariance are estimated during the M-step. {This avoids the need to handle the joint speech--noise posterior, but it also means that the noise signal itself is never explicitly reconstructed during inference.} We argue that optimizing the noise parameters in the M-step, without explicitly reconstructing the noisy observation as the sum of the estimated clean speech and acoustic noise, may lead either to poor data consistency or to well-verified data consistency but poor speech estimates due to inaccurate acoustic noise modeling. Besides, we hypothesize that modeling the acoustic noise with a more powerful model than NMF might improve SE.

In light of these findings, we propose new unsupervised diffusion-based SE frameworks that further reduce the performance gap with supervised approaches. Our contributions are threefold. \textbf{Firstly}, we present a comprehensive and unified description of previous work on unsupervised diffusion-based SE before introducing our new approaches, which build upon this unified formulation. \textbf{Secondly}, we go beyond \cite{ayilo2024diffavse}, which models the acoustic noise as a Gaussian distribution with a covariance structured by NMF, and propose to explicitly sample the acoustic noise from its posterior distribution. This entails treating the acoustic noise, similarly to the speech signal, as a latent variable within the observed noisy speech. {The main technical difficulty is that this leads to an intractable joint posterior over speech and noise. We address this by deriving a practical approximate EM inference scheme, in which the E-step uses Gibbs sampling to alternate between diffusion-based posterior sampling of speech and an explicit posterior update of noise, while the M-step updates the NMF parameters.} \textbf{Thirdly}, we propose to use a pre-trained diffusion-based prior model for noise, analogous to the approach used for speech, instead of the NMF-based model. Specifically, we present two algorithmic variants: one that does not treat the acoustic noise as a latent variable, and another that does, performing alternating diffusion-based posterior sampling to estimate both the clean speech and the acoustic noise. Contrary to \cite{chung2023parallel,11464305,mariani2024multisource}, which use separate diffusion models for the independent components of the mixture, our diffusion-based priors for speech and noise are trained jointly, thus reducing the number of models that must be trained. Moreover, the likelihood approximation technique used in the posterior inference of our approach is much simpler and lighter.

Experiments on the WSJ0-QUT \cite{leglaive2020recurrent} and VoiceBank-DEMAND \cite{botinhao2016investigating} datasets demonstrate that modeling noise as a latent variable, whose prior is either a Gaussian with an NMF-based covariance structure or a diffusion-based prior, improves the SE performance. Under mismatched conditions, the framework with an NMF-based noise covariance and explicit latent noise modeling outperforms its diffusion-based counterparts and exhibits less performance degradation than supervised baselines when evaluated on unseen data.

We structure the remainder of the paper as follows. 
In Sections~\ref{sec:diffusion_revision} and~\ref{sec:speech_diff_implicit_nmf_noise}, we review the fundamentals of score-based diffusion models (training and prior sampling), as well as unsupervised SE methods~\cite{nortier2023unsupervised, ayilo2024diffavse, sadeghi2025posterior} that combine diffusion models with NMF. Section~\ref{sec:speech_diff_explicit_nmf_noise} introduces our approach, which explicitly considers acoustic noise as a latent variable and performs SE using diffusion models with NMF. Section~\ref{sec:paradiffuse} presents our framework that uses diffusion-based prior models for both speech and noise data. The experimental setup and results are provided in Section~\ref{sec:exp}. Finally, we conclude in Section~\ref{sec:conc}.




\section{{Audio generative modeling with diffusion}}\label{sec:diffusion_revision}

In this section, we review the score-based generative diffusion model, which forms the basis for both the previously proposed unsupervised SE methods and the new approaches introduced in this work. We model audio in the complex-valued \ac{STFT} domain. For simplicity, a 2D \ac{STFT} array with $F$ frequency bins and $L$ time frames is represented by a flattened 1D vector $\vect{a} \in \mathbb{C}^{FL}$. We use $\vect{a}$ to denote a generic audio signal (either speech or background noise), so that all equations apply to both cases. When needed, we write $\vect{s}$ and $\vect{n}$ specifically for speech and noise, respectively.

\subsection{Diffusion modeling framework}\label{sec:diffusion_model_train}

Given a training set of audio data, the goal of a score-based diffusion model is to approximate the underlying data distribution $p_{\text{data}}(\vect{a})$. This line of work builds on score matching, which seeks to approximate the (generally intractable) score function $\nabla_{\vect{a}}\log p(\vect{a})$ so that new samples can be generated using this estimated score. In particular, denoising score matching~\cite{vincent2011connection} learns a neural network $\vect{S}_{\theta}$ to predict the gradient of the log-density of data that have been corrupted by additive Gaussian noise. \cite{song2019generative} {generalizes} this idea by introducing multiple noise scales, and \cite{song2021scorebased} further {extends} it by defining a continuous-time corruption process via a \ac{SDE} that gradually transforms data samples into Gaussian noise. This continuous-time corruption process is defined by the \textit{forward} \ac{SDE}:
\begin{equation}\label{eqn:sde-fwd}
    \textrm{d}\vect{a}_t = \vect{f}(\vect{a}_t) \textrm{d}t + g(t) \textrm{d}\vect{w},
\end{equation}
where $\vect{w}$ denotes a standard Wiener process, $\vect{f}$ is the drift coefficient, and $g(t)$ is the diffusion coefficient controlling the noise scale. We set $\vect{f}(\vect{a}_t) = -\gamma \vect{a}_t$, where $\gamma$ is a constant parameter.

The perturbation kernel associated with this \ac{SDE} allows us to sample $\vect{a}_t$ directly from a clean sample $\vect{a}$:
\begin{equation}\label{eqn:p0t-richter}
    p_{0t}(\vect{a}_t | \vect{a}) = \mathcal{N}_{\mathbb{C}}(\delta_t \vect{a}, \sigma(t)^2 \vect{I}),
\end{equation}
where $\delta_t = \exp(-\gamma t)$ and $\sigma(t)^2$ is derived from the \ac{SDE}.

To learn the score function, a neural network $\vect{S}_{\theta}(\vect{a}_t, t)$ is trained using the following objective:
\begin{equation}\label{eqn:train-obj}
    \theta^* = \argmin_{\theta} \mathbb{E}_{t, \vect{a}, \bs{\zeta}} 
    \Big[\| \sigma(t)\vect{S}_{\theta}(\vect{a}_t, t) + \bs{\zeta} \|_2^2\Big],
\end{equation}
where $\bs{\zeta} \sim \mathcal{N}_{\mathbb{C}}(\bs{0}, \vect{I})$ is the complex-valued zero-mean Gaussian noise used to corrupt the input data. Once trained, the score network can be used to generate new data samples via the \textit{reverse} \ac{SDE}, which under mild regularity assumptions, is given by~\cite{song2021scorebased}:
\begin{equation}\label{eqn:rev-sde_bis}
    \textrm{d}\vect{a}_t = \big[\vect{f}(\vect{a}_t) - g(t)^2 \nabla_{\vect{a}_t} \log p_t(\vect{a}_t)\big] \textrm{d}t + g(t) \textrm{d}\overline{\vect{w}},
\end{equation}
where $\overline{\vect{w}}$ is a standard Wiener process running backward in time, and $\textrm{d}t$ is a negative time increment.

This reverse \ac{SDE} progressively transforms a Gaussian noise sample into a clean data sample. Since the score $\nabla_{\vect{a}_t} \log p_t(\vect{a}_t)$ is unknown, it is approximated by the learned score network $\vect{S}_{\theta}$. Sampling is then performed by numerically solving the reverse \ac{SDE}. In our work, we adopt the \ac{PC} sampler from~\cite{song2021scorebased,richter2023speech}, which is described in the next section.


\subsection{{Diffusion prior sampling formulation}}\label{sec:prior_sampling}
The predictor step of the \ac{PC} sampler uses the Euler–Maruyama method to numerically integrate the reverse \ac{SDE}~\eqref{eqn:rev-sde_bis} in discrete time. This yields a finite sequence of latent variables $\{ \vect{a}_i \}_{i=0}^{N}$ forming a Markov chain
$\vect{a}_0 \rightarrow \vect{a}_1 \rightarrow \ldots \rightarrow \vect{a}_{N}$, whose joint distribution factorizes as
\begin{equation}\label{eq:joint_prior}
    p(\vect{a}_0,\ldots,\vect{a}_{N}) = p(\vect{a}_{N}) \prod_{i=1}^N p(\vect{a}_{i-1} \mid \vect{a}_{i}).
\end{equation}
Discretizing \eqref{eqn:rev-sde_bis} and inserting the score model leads to:
\begin{equation}\label{eq:disc-rev_logp}
  {\vect{a}}_{{i-1}} = {\vect{a}}_{i} - \vect{f}_i\Delta{\tau} + g_{\tau_i}^2 \vect{S}_{\theta^*}(\vect{a}_i, \tau_i)\Delta{\tau}+  g_{\tau_i}\sqrt{\Delta{\tau}}\bs{\zeta}
\end{equation} 
with $\vect{a}_{N}\sim \mathcal{N}_{\mathbb{C}}(\bs{0}, \Ib)$. Here, $\{ \tau_1,\ldots,\tau_N\} = \{i.T/N\}_{i=1}^{N}$ is an equally spaced sequence of time steps in $[0,T]$ with a discretization width of $\Delta{\tau}=T/N$, $\bs{\zeta}\sim\mathcal{N}_{\mathbb{C}}(\bs{0}, \Ib)$, and $\vect{f}_i=-\gamma \vect{a}_i$. Applying a similar discretization to the forward \ac{SDE} in \eqref{eqn:sde-fwd}, we obtain the \text{perturbation kernel}:
\begin{equation}\label{eq:ai_a0}
    p(\vect{a}_i | \vect{a}_0) = \mathcal{N}_{\mathbb{C}}(\delta_i \vect{a}_0, \sigma_{\tau_{i}}^2\vect{I}), \quad i\ge 1, \quad \delta_i = \exp(-\gamma \tau_{i}). 
\end{equation}

The predictor step given by~\eqref{eq:disc-rev_logp} can be refined by first applying a corrector step to its input $\vect{a}_i$. Following~\cite{song2021scorebased}, we use Langevin Markov Chain Monte Carlo (MCMC) sampling as the corrector. This consists of updating $\vect{a}_i$ via a gradient-ascent step on the log-density, augmented with Gaussian noise, so as to move towards regions of higher probability, where the gradient term is the estimated score function. Overall, the prior sampling is then characterized by the following backward transition distribution:
\begin{equation}\label{eq:cond_prior}
    p(\vect{a}_{{i-1}}|\vect{a}_{i})=\mathcal{N}_{\mathbb{C}}\Big(\bs{\mu}^{\mathsf{back}}(\vect{a}_{i}), \bs{\Sigma}^{\mathsf{back}}_{i}\Big),
\end{equation}
with
\begin{equation}\label{eq:mu_sigma_update_prior}
  \left\{
  \mathopen{}\begin{array}{l}
    \bs{\mu}^{\mathsf{back}}(\vect{a}_{i}) =\vect{h}({\vect{a}}_{i}) - \vect{f}_{i}\Delta{\tau} + g_{\tau_i}^2 \vect{S}_{\theta^*}(\vect{h}({\vect{a}}_{i}), \tau_i)\Delta{\tau} \\[2mm]
    \bs{\Sigma}^{\mathsf{back}}_{i} = g_{\tau_i}^2\Delta{\tau}\Ib.
  \end{array}
  \right.
\end{equation}
where $\vect{h}({\vect{a}}_{i})$ is the result of one-step Langevin MCMC: 
\begin{equation}\label{eq:update_h}
    \vect{h}({\vect{a}}_{i}) = \vect{a}_{i} + \epsilon_{\tau_i} {\vect{S}_{\theta^*}(\vect{a}_{i}, \tau_i)} + \sqrt{2\epsilon_{\tau_i}}\bs{\zeta},\quad \bs{\zeta}\sim\normalc.
\end{equation}
Note that $\epsilon_{\tau_i} = (\sigma_{\tau_i} \cdot r)^2$ denotes the Langevin step size ($r>0$) \cite{song2022solving}. Starting from Gaussian noise, i.e., $\vect{a}_{N}\sim \mathcal{N}_{\mathbb{C}}(\bs{0}, \Ib)$, the above conditional sampling procedure is iterated to ultimately generate an audio signal $\vect{a}_0$ from the learned distribution. 
\subsection{{Tweedie's formula}}\label{sec:tweedie_paragraph}
Numerous works use Tweedie’s formula~\cite{efron2011tweedie} to solve inverse problems with diffusion models. Tweedie’s formula provides an expression for the posterior mean of the intractable distribution $p(\vect{a}_0 | \vect{a}_i)$. Given the Gaussian conditional density $p(\vect{a}_i | \vect{a}_0)$ in~\eqref{eq:ai_a0}, the posterior mean $\E_{p_{i0}(\vect{a}_0 | \vect{a}_i)}[\vect{a}_0]$ can be approximated as
\begin{equation}\label{eq:tweedie}
    \hat{\vect{a}}_{0,i}
    = \E_{p_{i0}(\vect{a}_0 | \vect{a}_i)}[\vect{a}_0]
    \approx \frac{\vect{a}_i + \sigma_{\tau_i}^2 \vect{S}_{\theta^*}(\vect{a}_i, \tau_i)}{\delta_{i}}.
\end{equation}
This expectation corresponds to a Gaussian denoising operation applied to the perturbed signal $\vect{a}_i$, yielding an estimate of $\vect{a}_0$ at every reverse iteration.

\section{Prior work: {Diffusion-based speech and NMF-based noise priors with implicit noise modeling}}\label{sec:speech_diff_implicit_nmf_noise}
\subsection{Problem formulation}
To perform SE, we assume the additive mixture model
\begin{equation}\label{eq:obs_model_implicit}
    \vect{x} = \vect{s} + \vect{n},
\end{equation}
where $\vect{x}$, $\vect{s}$, and $\vect{n}$ denote the \ac{STFT} arrays of noisy speech, clean speech, and background noise, respectively. The additive noise is modeled as
$\vect{n} \sim \mathcal{N}_{\mathbb{C}}(\bs{0}, \diag(\vect{v}_{\phi}))$, where
$\vect{v}_{\phi} = \text{vec}(\vect{W}\vect{H})$,
$\vect{W}$ and $\vect{H}$ are low-rank, non-negative matrices, and $\text{vec}(\cdot)$ denotes the vectorization operator. Concretely, $\vect{W}$ contains spectral templates, and $\vect{H}$ encodes their time-varying activations~\cite{vincent2018audio}. The product $\vect{W}\vect{H}$ is intended to approximate the true noise spectrogram. {This assumption follows the Gaussian Composite Model used in NMF-based audio source modeling, where complex STFT coefficients are modeled as independent proper complex Gaussian variables with variances given by the source power spectrogram~\cite{fevotte2009nonnegative,vincent2018audio}. Here, this model is used only for background noise.}

In the unsupervised SE setting considered in prior work, the goal is to estimate the clean speech $\vect{s}$ together with the noise parameters $\phi = \{\vect{W}, \vect{H}\}$. This amounts to maximizing the observed-data log-likelihood given by:
\begin{equation}\label{eq:observed_data_ll}
    \mathcal{L}(\phi;\vect{x})=\log p_\phi(\vect{x})=\log \int p_\phi(\vect{x}, \vect{s}) \textrm{d} \vect{s},
\end{equation}
which is intractable to compute. Instead, an EM procedure is followed by iteratively optimizing the following expected complete-data log-likelihood over $\phi$:
\begin{equation}\label{eq:complete_data_log}
    \mathcal{Q}(\phi;\phi_c) = \E_{p_{\phi_c}(\vect{s}|\vect{x})}[\log p_\phi(\vect{x}, \vect{s})],
\end{equation}
where $\phi_c$ denotes the current estimate of $\phi$. In the E-step, the intractable expectation is approximated via Monte Carlo, by drawing samples from the posterior $p_{\phi_c}(\vect{s} | \vect{x})$ given the current parameter estimate $\phi_c$. More precisely, we have:
\begin{equation}
    \widehat{\mathcal{Q}}(\phi;\phi_c) = \frac{1}{B} \sum_{b=1}^{B} \log p_\phi(\mathbf{x}, \mathbf{s}^b),
\end{equation}
where $\mathbf{s}^b \sim p_{\phi_c}(\mathbf{s} | \mathbf{x})$ and $B$ is the number of posterior samples. The M-step corresponds to the optimization problem below:
\begin{align}
    {\phi^{*}} & \approx \underset{{\phi} \geq 0}{\textrm{argmax}}\,[ \widehat{\mathcal{Q}}(\phi;\phi_c)]\\ &= \underset{{\phi} \geq 0}{\textrm{argmax}}\frac{1}{B} \sum_{b=1}^{B}\sum_{j} \frac{(\vect{x}-\mathbf{s}^{b})^{H}_j(\vect{x}-\mathbf{s}^{b})_j}{\vect{v}_{\phi, j}} + \log \vect{v}_{\phi, j},
\end{align}
which is solved using the multiplicative update rules \cite{fevotte2009nonnegative}. In the above formula, the subscript $j$ denotes the $j$\textsuperscript{th} entries of the associated variables, and $H$ is the conjugate transpose.

Recall that the goal of SE is to recover an accurate estimate of the clean speech signal. In the EM framework, this objective is achieved by sampling from the posterior in the E-step. The more accurate the speech prior and the noise parameters $\phi$ are, the more reliable the posterior samples of $\vect{s}$ become. What remains is to clarify how to sample from this posterior using the diffusion model. Depending on the chosen construction, this leads to different E-step algorithms.
\begin{figure*}[t!]
 \centering
 \includegraphics[width=0.85\linewidth]{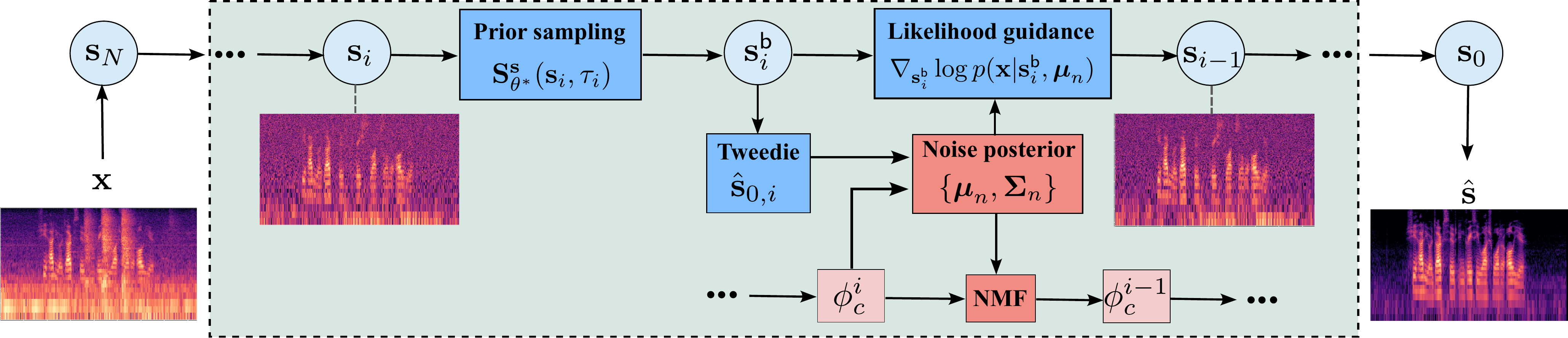}
 \caption{Schematic diagram of the \texttt{DiffUSEEN} algorithm.}
 \label{fig:diffuseen}
\end{figure*}

As discussed in Section~\ref{sec:prior_sampling}, unconditional sampling from the clean-speech distribution is performed by estimating the score function $\nabla_{\vect{s}_i} \log p_i(\vect{s}_i)$ and integrating the reverse SDE~\eqref{eq:disc-rev_logp}. In a similar spirit, a common way to sample from the posterior $p_{\phi_c}(\vect{s} | \vect{x})$, given the current noise parameter $\phi_c$ obtained at the previous M-step, is to approximate the posterior score. By Bayes’ rule,
$p_{\phi_c}(\vect{s} | \vect{x}) \propto p_{\phi_c}(\vect{x} | \vect{s})\, p(\vect{s})$,
so the corresponding score decomposes as a sum of a likelihood term and the prior score:
\begin{equation}
    \nabla_{\vect{s}_i} \log p_{\phi_c}(\vect{s}_i|\vect{x}) = \nabla_{\vect{s}_i} \log p_{\phi_c}(\vect{x}|\vect{s}_i)+ \nabla_{\vect{s}_i} \log p_i(\vect{s}_i). 
\end{equation}
Replacing the second term with the learned score model, the associated reverse \ac{SDE} used to sample from the posterior at the current EM iteration can then be written as:
\begin{equation}\label{eq:disc-rev-post-disc_logp}
\begin{split}
    {\vect{s}}_{{i-1}} = {\vect{s}}_{i} - \vect{f}_i\Delta{\tau} + g_{\tau_i}^2 [\nabla_{\vect{s}_i} \log p_{\phi_c}(\vect{x} | \vect{s}_i) + \vect{S}^{\vect{s}}_{\theta^*}(\vect{s}_{i}, \tau_i)]\Delta{\tau}\\+  g_{\tau_i}\sqrt{\Delta{\tau}}\bs{\zeta}.
\end{split}
\end{equation}
Regarding the likelihood term $p_{\phi_c}(\vect{x}| \vect{s}_i)$, marginalizing over $\vect{s}$ gives
\begin{equation}\label{eq:likelihood_x_phi}
    p_{\phi_c}(\vect{x}| \vect{s}_i)
    = \int p_{\phi_c}(\vect{x}| \vect{s}) \, p(\vect{s}| \vect{s}_i)\, \mathrm{d}\vect{s},
\end{equation}
which is intractable due to
$p(\vect{s}| \vect{s}_i)$.
A large body of work has proposed approximations of this likelihood in the context of inverse problems with diffusion models~\cite{daras2024survey}. In the following subsections, we briefly review how~\cite{nortier2023unsupervised, ayilo2024diffavse, sadeghi2025posterior} approximate $p_{\phi_c}(\vect{x}| \vect{s}_i)$ and perform the posterior sampling required in the E-step.
\subsection{\texttt{UDiffSE}} 
To simplify the likelihood computation at inference, a non-informative prior assumption was proposed by \cite{meng2024diffusion}, which leads to $p(\vect{s}|\vect{s}_i) \propto p(\vect{s}_i|\vect{s}) p(\vect{s}) \propto p(\vect{s}_i|\vect{s})$. \texttt{UDiffSE} \cite{nortier2023unsupervised} leverages this assumption, which combined with \eqref{eq:ai_a0} leads to: 
\begin{equation}\label{eq:uninformative}   
p(\vect{s}|\vect{s}_i) \approx \tilde{p}(\vect{s}|\vect{s}_i)=\mathcal{N}_{\mathbb{C}}\Bigl(\frac{\vect{s}_i}{{\delta}_i}, \frac{\sigma_{\tau_i}^2}{{\delta}_i^2}\vect{I} \Bigr).
\end{equation}
This results in an approximate likelihood, referred to as the \textit{noise-perturbed pseudo-likelihood}, defined as:
\begin{equation}\label{eq:likelihood_udiffse}   
\tilde{p}_{\phi_c}(\vect{x}|\vect{s}_i) = \mathcal{N}_{\mathbb{C}}\Big(\frac{\vect{s}_i}{{\delta}_i}, \mathbf{J}_{i, \phi_c}\Big),
\end{equation}
where $\mathbf{J}_{i,\phi_c} = \frac{\sigma_{\tau_i}^2}{{\delta}_{i}^2} \vect{I} + \diag(\vect{v}_{\phi_c})$. Therefore, we have
\begin{equation}\label{eq:likelihood_score_imp}
    \nabla_{\vect{s}_i}{\log\,} \tilde{p}_{\phi_c}(\vect{x}|\vect{s}_i) = \dfrac{1}{{\delta_i}}\,\mathbf{J}^{-1}_{i,\phi_c}\,\Bigl(\vect{x}-\dfrac{{\vect{s}}_{i} }{\delta_i}\Bigr). 
\end{equation}
{Note that $\mathbf{J}$ is a diagonal matrix and that all its diagonal entries are non-zero, so it can be accurately inverted.}

\noindent A scaling factor $\lambda_i$ is introduced to balance the contributions of the prior score and the likelihood score in the posterior sampling step formulated below:
\begin{equation}\label{eq:disc-rev-post_disc_logp_approx}
\begin{split}
    {\vect{s}}_{{i-1}} = {\vect{s}}_{i} - \vect{f}_i\Delta{\tau} + g_{\tau_i}^2 \Big[\lambda_i\nabla_{\vect{s}_i} \log \tilde{p}_{\phi_c}(\vect{x} | \vect{s}_i) + \\ \vect{S}^{\vect{s}}_{\theta^*}(\vect{s}_i, \tau_i)\Big]\Delta{\tau}+  g_{\tau_i}\sqrt{\Delta{\tau}}\bs{\zeta}.
\end{split}    
\end{equation}
At each E-step, the reverse SDE iterations in~\eqref{eq:disc-rev-post_disc_logp_approx} are performed, starting from Gaussian noise. The resulting samples are then used in the M-step to update the parameter $\phi$, and the EM steps are repeated until convergence (typically within five iterations) \cite{nortier2023unsupervised}.
\subsection{\texttt{UDiffSE+}} 
To reduce the inference time of \texttt{UDiffSE}, \cite{ayilo2024diffavse} propose interleaving the M-step with the reverse iterations in~\eqref{eq:disc-rev-post_disc_logp_approx}, and using Tweedie’s formula~\eqref{eq:tweedie} to obtain the clean-speech estimate required to update $\phi$. In this way, the noise parameters are updated progressively during the reverse pass, rather than only after it is complete, which reduces the overall number of reverse passes to a single round.

\subsection{{\texttt{DEPSE-IL} and \texttt{DEPSE-TL}}}
Instead of explicitly incorporating the likelihood gradient term 
$\nabla_{\vect{s}_i} \log p_{\phi_c}(\vect{x} | \vect{s}_i)$
to guide the generation of clean speech from noisy observations, \cite{sadeghi2025posterior} propose to model directly the transition distribution
\[
p_{\phi_c}(\vect{s}_{i-1} | \vect{s}_{i}, \vect{x})
\propto p_{\phi_c}(\vect{x} | \vect{s}_{i-1})\, p(\vect{s}_{i-1} | \vect{s}_{i}),
\]
where $p(\vect{s}_{i-1} | \vect{s}_{i})$ is the prior transition used in~\eqref{eq:cond_prior}.  
This approach, called \texttt{DEPSE-IL}, removes the need for the scaling factor $\lambda_i$, but still requires an approximation of the likelihood term.

In addition, \cite{sadeghi2025posterior} introduce an alternative strategy, \texttt{DEPSE-TL}, which applies the forward diffusion process to the noisy speech (using the discretized form of~\eqref{eqn:p0t-richter} applied to $\vect{x}$).  
This yields a modified transition distribution
\[
p_{\phi_c}(\vect{s}_{i-1} | \vect{s}_{i}, \vect{x}_{i-1})
\propto p_{\phi_c}(\vect{x}_{i-1} | \vect{s}_{i-1})\, p(\vect{s}_{i-1} | \vect{s}_{i}),
\]
for which the likelihood term $p_{\phi_c}(\vect{x}_{i-1} | \vect{s}_{i-1})$ is fully tractable.
\section{{Diffusion-based speech and NMF-based noise priors with explicit noise modeling}}\label{sec:speech_diff_explicit_nmf_noise}
In the previous sections, we reviewed unsupervised SE methods that rely on a diffusion-based speech prior and NMF-based noise modeling, where only the speech $\vect{s}$ is treated as a latent variable. The noise $\vect{n}$ affects inference only through its NMF-structured covariance in the likelihood term $p(\vect{x} | \vect{s})$. {This simplifies inference, but the noise signal itself is neither reconstructed nor sampled from its posterior distribution. As a result, mixture consistency is enforced only implicitly through an NMF-weighted likelihood term, which may be poorly conditioned when the NMF noise model is inaccurate.}

To address this limitation, we introduce \texttt{DiffUSEEN} (Diffusion-based Unsupervised Speech Enhancement with Explicit Noise Modeling), illustrated in Fig.~\ref{fig:diffuseen}. {The key idea is to treat both $\vect{s}$ and $\vect{n}$ as latent variables and to make this formulation tractable through an approximate EM procedure: the E-step alternates between diffusion-based posterior sampling of speech and an explicit posterior update of noise, while the M-step updates the NMF parameters. This enables explicit speech--noise reconstruction and better enforces the mixture constraint $\vect{x}\approx \hat{\vect{s}}+\hat{\vect{n}}$.} This formulation also prepares the ground for Section~\ref{sec:paradiffuse}, where we replace NMF-based noise modeling with a learned diffusion prior.

In the proposed approach, we explicitly model the noise $\vect{n}$ as a latent variable with prior $\vect{n} \sim \mathcal{N}_{\mathbb{C}}(\bs{0}, \diag(\vect{v}_{\phi}))$, parameterized by NMF as before. {We further introduce a small Gaussian residual term, which yields a tractable likelihood and turns the hard mixture constraint into a soft consistency term.} Specifically, we assume:
\begin{equation}\label{eq:obs_model_explicit}
    \vect{x} = \vect{s} + \vect{n} + \vect{r},
\end{equation}
where $\vect{r} \sim \mathcal{N}_{\mathbb{C}}(\bs{0}, \sigma_r^2 \vect{I})$, and $\sigma_r^2$ is a known, small constant.

The noise parameters $\phi$ are estimated by maximizing the observed-data log-likelihood $\log p_\phi(\vect{x})$, similarly to \eqref{eq:observed_data_ll}, which, however, leads to an intractable expression. To address this, we follow the EM procedure and instead optimize the expected complete-data log-likelihood over $\phi$:
\begin{equation}\label{eq:complete_data_log_en}
    \mathcal{Q}(\phi;\phi_c) = \E_{p_{\phi_c}(\vect{s}, \vect{n}|\vect{x})}[\log p_\phi(\vect{x}, \vect{s}, \vect{n})],
\end{equation}
where $\phi_c$ denotes the current estimate of $\phi$.

To approximate the expectation in~\eqref{eq:complete_data_log_en} via Monte Carlo, we sample from the joint posterior $p_{\phi_c}(\vect{s}, \vect{n} | \vect{x})$. We do so using Gibbs sampling, decomposing the joint draw into two conditional updates: first, we sample $\vect{s}$ from $p(\vect{s} | \vect{x}, \vect{n})$ given the current estimate of $\vect{n}$; then, we sample $\vect{n}$ from $p_{\phi_c}(\vect{n} | \vect{x}, \vect{s})$ using the newly updated $\vect{s}$.
 Mathematically,
\begin{subequations}
    \begin{empheq}[left={\empheqlbrace}]{align}
        \vect{s} &\sim p(\vect{s}|\vect{x}, \vect{n}), \label{eq:s_update_gibbs}\\
        \vect{n} &\sim p_{\phi_c}(\vect{n}|\vect{x}, \vect{s}). \label{eq:n_update_gibbs}
    \end{empheq}
\end{subequations}
{In practice, the exact speech conditional distribution is intractable and is approximated through diffusion-based posterior sampling, whereas the noise conditional admits a tractable Gaussian form under the likelihood approximation introduced below.}
\subsubsection{{E-step for sampling $\vect{s}$}} To sample $\vect{s}$ in \eqref{eq:s_update_gibbs}, we follow the procedure of \texttt{UDiffSE+}~\cite{ayilo2024diffavse} by replacing $\nabla_{\vect{s}_i} \log p_i(\vect{s}_i)$ in \eqref{eq:disc-rev_logp} with the following posterior-based score term:
\begin{equation}
\label{eq:log_p_s_dec_v2}
    \nabla_{\vect{s}_i} \log p(\vect{s}_i|\vect{x}, \vect{n}) 
    = \nabla_{\vect{s}_i} \log p(\vect{x}|\vect{s}_i, \vect{n}) 
    + \nabla_{\vect{s}_i} \log p_i(\vect{s}_i).
\end{equation}
We also need to approximate the intractable likelihood $p(\vect{x}|\vect{s}_i, \vect{n})$:
\begin{align}\label{eq:likelihood_fudiffse_v2}
    p(\vect{x}|\vect{s}_i, \vect{n}) 
    &= \int p(\vect{x},\vect{s}_0|\vect{s}_i, \vect{n})\textrm{d}\vect{s}_0 \nonumber\\
    &= \int p(\vect{x}|\vect{s}_0, \vect{n}) \, p(\vect{s}_0|\vect{s}_i)\,\textrm{d}\vect{s}_0.
\end{align}
Here, $p(\vect{x}| \vect{s}_0, \vect{n})=\mathcal{N}_{\mathbb{C}}(\vect{s}_0+\vect{n}, \sigma_r^2 \vect{I})$.  
Assuming an uninformative prior $p(\vect{s}_0)$ and using \eqref{eq:ai_a0}, we obtain the following Gaussian approximation, {which makes the likelihood tractable but may introduce approximation bias}:
\begin{equation}\label{eq:approx_likelihood}
    \tilde{p}(\vect{x}|\vect{s}_i, \vect{n}) 
    = \mathcal{N}_{\mathbb{C}}\Big(
        \frac{\vect{s}_i}{{\delta}_i}+\vect{n},
        \sigma_x^2\vect{I}
    \Big),
\end{equation}
where $\sigma_x^2={\sigma^2_{\tau_i}}/{{\delta}_i^2}+\sigma_r^2$. Compared to \eqref{eq:likelihood_udiffse}, 
the covariance matrix here is diagonal, and the mixture consistency is explicitly taken into account. The likelihood score term in \eqref{eq:log_p_s_dec_v2} is then approximated as:
\begin{equation}\label{eq:noise_likelihood}
    \nabla_{\vect{s}_i} \log \tilde{p} (\vect{x}|\vect{s}_i, \vect{n}) 
    = \frac{1}{{\delta}_i \sigma_x^2} 
      \Big(\vect{x}-\Big(\frac{\vect{s}_i}{{\delta}_i}+\vect{n}\Big)\Big).
\end{equation}
Finally, the clean speech samples are drawn using the reverse \ac{SDE}:
\begin{equation}\label{eq:disc-rev-post_disc_logp_approx_fudiffse_v2} \begin{split} {\vect{s}}_{{i-1}} = {\vect{s}}_{i} - \vect{f}_i\Delta{\tau} + g_{\tau_i}^2 \Big[\lambda_i\nabla_{\vect{s}_i} \log \tilde{p}(\vect{x} | \vect{s}_i, \vect{n}) + \\ \vect{S}^{\vect{s}}_{\theta^*}(\vect{s}_i, \tau_i)\Big]\Delta{\tau}+ g_{\tau_i}\sqrt{\Delta{\tau}}\bs{\zeta}. \end{split} \end{equation}
In practice, $\lambda_i$ is set using a scheduler function. Moreover, the reverse \ac{SDE} \eqref{eq:disc-rev-post_disc_logp_approx_fudiffse_v2} is numerically solved by first running a prior sampling step, as done in \eqref{eq:cond_prior}-\eqref{eq:update_h}, followed by a data-consistency update (via the posterior score), to take into account $\vect{x}$. 

\subsubsection{{E-step for sampling $\vect{n}$}}   
To sample $\vect{n}$, we need to compute the posterior in \eqref{eq:n_update_gibbs} within the reverse diffusion steps, that is
\begin{equation}
      p_{\phi_c}(\vect{n}| \vect{x}, \vect{s}_i)\propto p(\vect{x}| \vect{s}_i, \vect{n})\,p_{\phi_c}(\vect{n}).
\end{equation}
We approximate the intractable likelihood term using \eqref{eq:approx_likelihood}, which yields the following Gaussian approximation to the noise posterior
\begin{equation}\label{eq:n_posterior}
    p_{\phi_c}(\vect{n}| \vect{x}, \vect{s}_i) \approx \mathcal{N}_{\mathbb{C}}(\boldsymbol{\mu}_n, \boldsymbol{\Sigma}_n),
\end{equation} 
where
\begin{subequations}
\label{eq:mu_sigma_update_noise}
    \begin{empheq}[left={\empheqlbrace}]{align}
        \boldsymbol{\mu}_n &=\frac{\vect{v}_{\phi_c}}{\sigma_x^2 + \vect{v}_{\phi_c}}{(\vect{x}-\frac{\vect{s}_i}{{\delta}_i})}, \label{eq:n_mean}\\
        \boldsymbol{\Sigma}_n &= \frac{\sigma_x^2\vect{v}_{\phi_c}}{\sigma_x^2 + \vect{v}_{\phi_c}}, \label{eq:n_var}
    \end{empheq}
\end{subequations}
Here, divisions and multiplications are to be interpreted element-wise.  
When computing \eqref{eq:n_mean}, we replace $\vect{s}_i/\delta_i$ with the Tweedie-based speech estimate $\hat{\vect{s}}_{0,i}$ given in \eqref{eq:tweedie}, as it performs much better in our experiments.
The posterior mean $\boldsymbol{\mu}_n$ is used as the current estimate of $\vect{n}$ in \eqref{eq:disc-rev-post_disc_logp_approx_fudiffse_v2}.  

\subsubsection{{M-step}}
As stated earlier, the M-step maximizes~\eqref{eq:complete_data_log_en} with respect to $\phi$, which can be rewritten as
\begin{equation}
    \max_{\phi}\ \E_{p_{\phi_c}(\vect{s}, \vect{n}|\vect{x})} [\log p_\phi(\vect{n})].
\end{equation}
By factorizing
$p_{\phi_c}(\vect{s}, \vect{n}|\vect{x}) = p_{\phi_c}(\vect{n}|\vect{x}, \vect{s})\, p(\vect{s}|\vect{x})$
and noting that $p(\vect{s}|\vect{x})$ does not depend on $\phi$, this simplifies to
\begin{equation}
    \max_{\phi}\ \E_{p_{\phi_c}(\vect{n}|\vect{x}, \vect{s})} [\log p_\phi(\vect{n})].
\end{equation}
Substituting the prior distribution of $\vect{n}$, the above problem becomes
\begin{equation}
    \min_{\phi}\ \E_{p_{\phi_c}(\vect{n}|\vect{x}, \vect{s})} \Big[\sum_j\frac{| n_{j}|^2}{v_{\phi, j}}+\log v_{\phi, j}\Big], 
\end{equation}
which, by incorporating the noise posterior \eqref{eq:n_posterior}, results in
\begin{equation}
    \min_{\phi}\ \sum_{{j}} \frac{| {\mu}_{n, j}|^2 + \Sigma_{n,j}}{v_{\phi, j}}+\log v_{\phi, j}, 
\end{equation}
where the subscript $j$ denotes the $j$\textsuperscript{th} entry of the associated variables. This is a standard NMF problem with the Itakura-Saito (IS) divergence loss, which can be solved using the same multiplicative update rules as in the previous algorithms. The overall algorithm iterates over all the steps discussed above and is summarized in Alg.~\ref{alg:fudiffsev2}.
\begin{figure*}[t]
 \centering
 \includegraphics[width=0.85\linewidth]{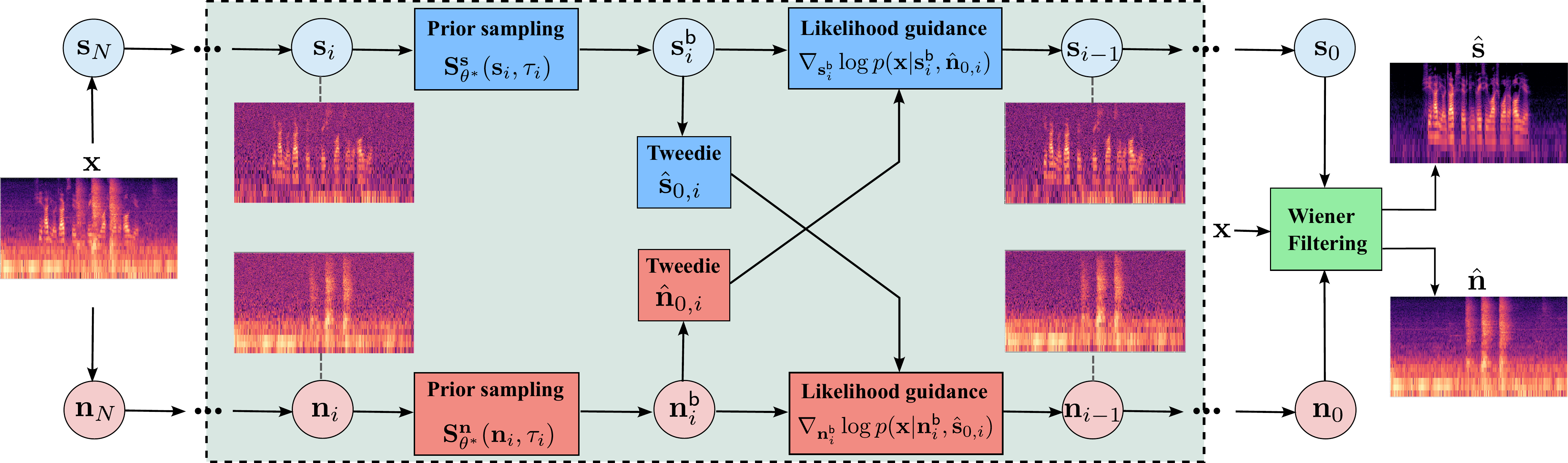}
 \caption{Schematic diagram of the ParaDiffUSE algorithm.}
 \label{fig:paradiffuse}
\end{figure*}
\begin{algorithm}[t!]
  \footnotesize
  \setstretch{1.3}
\caption{\texttt{DiffUSEEN}}\label{alg:fudiffsev2}
\begin{algorithmic}[1]
\Require $\vect{x}, N, \lambda, \sigma_r^2, \phi_c^{N}=\{\mathbf{W}_0,\mathbf{H}_0\}$
\State ${\vect{s}}_N \sim \mathcal{N}_{\mathbb{C}}(\vect{x}, \sigma^2_{\tau_N}\vect{I}), \Delta \tau \gets \frac{1}{N}$
\For{$i=N,\ldots, 1$}
  \State $\tau_i\gets i \Delta$, $\lambda_i \gets \textsc{scheduler}_\lambda(i )$
  \State \text{Compute} $\{ \bs{\mu}^{\mathsf{back}}(\vect{s}_{i}), \bs{\Sigma}^{\mathsf{back}}_i\}$ using \eqref{eq:mu_sigma_update_prior} \Comment{\textit{(Predictor-Corrector)}}
  \vspace{0.2cm}
  \State ${\vect{s}}^{\mathsf{b}}_{i} \gets \bs{\mu}^{\mathsf{back}}(\vect{s}_{i}) + \sqrt{\bs{\Sigma}^{\mathsf{back}}_i}\bs{\zeta}, \quad \bs{\zeta}\sim \normalc$
  \vspace{0.2cm}
  \State ${\hat{\mathbf{s}}}_{0,i}
        = \delta_{i}^{-1}\Big({\vect{s}}^{\mathsf{b}}_{i} + \sigma^2_{\tau_i} \vect{S}_{\theta^*}^{\vect{s}}({\vect{s}}^{\mathsf{b}}_{i},\tau_i)\Big)$ \Comment{\textit{Estimate of $\vect{s}_0$}}
  \vspace{0.2cm}
  \State Compute $\{ \boldsymbol{\mu}_n, \boldsymbol{\Sigma}_n \}$ via \eqref{eq:mu_sigma_update_noise} with $\vect{s}={{\hat{\mathbf{s}}}_{0,i}}$ \Comment{\textit{Noise posterior}}
  \vspace{0.2cm}

 \State $\vect{s}_{i-1} \gets {\vect{s}}^{\mathsf{b}}_{i}
        + {\lambda_i}{g_{i}^2}\,\nabla_{\vect{s}_i} \log \tilde{p}_{\phi_c^{i}} \!\left(\vect{x}\mid{\vect{s}}^{\mathsf{b}}_{i}, \boldsymbol{\mu}_n\right)\, \Delta{\tau}$


  \vspace{0.2cm}
  \State {$\phi_c^{i-1} = \textsc{NMF}(|\boldsymbol{\mu}_n|^2 + \boldsymbol{\Sigma}_n)$} \Comment{\textit{Parameters update}}
\EndFor
\State \textbf{Output:} $\hat{\vect{s}}={\vect{s}}_{0}$
\end{algorithmic}
\end{algorithm}

\section{{Diffusion-based speech and noise priors}}\label{sec:paradiffuse}
In the previous methods, noise was modeled by imposing an NMF structure on its covariance.  
Despite the success of this strategy, we hypothesize that performance can be improved with a more expressive noise model. Therefore, we propose to model the noise prior distribution with a diffusion model, as is done for clean speech. Furthermore, instead of learning two separate score models for speech and noise, we train a \emph{joint} score model conditioned on a label $\kappa$ indicating whether the input is speech ($\kappa = 1$) or noise ($\kappa = 0$).

In this framework, SE is addressed purely through diffusion-based posterior sampling. Since no noise-related parameter is optimized at inference, there is no M-step, in contrast to the previous EM-based approaches; only a posterior sampling procedure is required. We refer to this framework as \texttt{ParaDiffUSE} (Parallel Diffusion Model for Unsupervised Speech Enhancement), depicted in Fig.~\ref{fig:paradiffuse}. {It can be seen as a semi-supervised method rather than a fully unsupervised one, as it uses noise data in addition to clean speech data.} Within \texttt{ParaDiffUSE}, we propose two variants: \texttt{ParaDiffUSE-IN}, which \textit{implicitly} accounts for noise in the posterior sampling, and \texttt{ParaDiffUSE-EN}, which \textit{explicitly} samples from the posterior noise distribution. In both cases, we assume access to a pre-trained joint speech–noise score model, denoted
$\vect{S}_{\psi}(\vect{s}_t, t, \kappa)$. For brevity, we write
$\vect{S}_{\psi}^{\vect{s}}(\vect{s}_t, t) = \vect{S}_{\psi}(\vect{s}_t, t, 1)$
and
$\vect{S}_{\psi}^{\vect{n}}(\vect{n}_t, t) = \vect{S}_{\psi}(\vect{n}_t, t, 0)$.

Compared to \cite{11464305}, which uses separate models for speech and acoustic noise, our joint model can naturally be extended to multiple labels,
$\{\text{speech}, \text{noise}_1, \text{noise}_2, \ldots\}$,
which is useful when the noise type is known or can be estimated beforehand. A separate-model approach would then require training and maintaining several distinct noise models, which may be impractical. {Unlike supervised methods, \texttt{ParaDiffUSE}-type models do not require paired clean--noisy data, since separate speech and noise corpora are sufficient. This avoids relying on predefined mixtures and allows the noise branch to be fine-tuned on target-domain noise without constructing new noisy data.}
\begin{algorithm}[t!]
  \footnotesize
  \setstretch{1.3}  
\caption{\texttt{ParaDiffUSE-IN}}\label{alg:ParadiffuSEv1}
\begin{algorithmic}[1]
\Require $\vect{x}, N, \lambda$
    \State ${\vect{s}}_N \sim \mathcal{N}_{\mathbb{C}}(\vect{x}, \sigma^2_{\tau_N}\vect{I}), \Delta \tau \gets \frac{1}{N}$
    \For{$i=N,\ldots, 1$}
        \State $\tau_i\gets i \Delta$, $\lambda_i \gets \textsc{scheduler}_\lambda(i )$

        \vspace{0.2cm} 
        \State \text{Compute} $\{ \bs{\mu}^{\mathsf{back}}(\vect{s}_{i}), \bs{\Sigma}^{\mathsf{back}}_i\}$ using \eqref{eq:mu_sigma_update_prior} \Comment{\textit{(Predictor-Corrector)}}

        \vspace{0.2cm}
        \State ${\vect{s}}^{\mathsf{back}}_{i} \gets \bs{\mu}^{\mathsf{back}}(\vect{s}_{i}) + \sqrt{\bs{\Sigma}^{\mathsf{back}}_i}\bs{\zeta}, \quad \bs{\zeta}\sim \normalc$

        \vspace{0.2cm}  

        \State ${\vect{n}_i \gets \vect{x} - {\delta}_i^{-1} ({\vect{s}}^{\mathsf{back}}_{i}-\sigma_{\tau_i}\bs{\zeta}), \quad \bs{\zeta}\sim \normalc}$

        \vspace{0.2cm} 
        \State $\nabla_{\vect{s}^{\mathsf{back}}_i} \log \tilde{p}(\vect{x}|\vect{s}^{\mathsf{back}}_i) \gets -\delta_i^{-1}\vect{S}_{\psi^*}^{\vect{n}}(\vect{n}_i, \tau_i)$ \label{lst:line:gradient_ll_noise_paradiffusein}

        \vspace{0.2cm} 

            \State $\vect{s}_{i-1} \gets {\vect{s}}^{\mathsf{back}}_{i}  + {\lambda_i}{g_{i}^2} {\nabla_{\vect{s}^{\mathsf{back}}_i} \log \tilde{p}(\vect{x}|\vect{s}^{\mathsf{back}}_i)} \Delta{\tau}$ \Comment{\textcolor{darkgray}{$\vect{x}$-consistency}}
            \vspace{0.2cm}
    \EndFor
    \State \textbf{Output:} $\hat{\vect{s}}={\vect{s}}_{0}$
\end{algorithmic}
\end{algorithm}
\subsection{{Implicit noise modeling:} \texttt{ParaDiffUSE-IN}}
We first adopt an implicit noise-modeling framework, as in the previous methods~\cite{nortier2023unsupervised, ayilo2024diffavse, sadeghi2025posterior}, with the key difference that the noise prior is now modeled by a diffusion process instead of NMF. Using the observation model in~\eqref{eq:obs_model_implicit}, our goal is to recover the clean speech signal via posterior sampling, leveraging the pre-trained (joint) speech and noise diffusion model. To this end, we follow the \texttt{UDiffSE+} framework, in which the likelihood is written as
\begin{align}\label{eq:likelihood_paradiffuse_v1}
    p(\vect{x}|\vect{s}_{i}) = \int p(\vect{x},\vect{s}|\vect{s}_{i})\,\textrm{d}\vect{s}\nonumber &= \int p(\vect{x}|\vect{s})\,p(\vect{s}|\vect{s}_{i})\,\textrm{d}\vect{s}\nonumber\\
    &= \int p_n(\vect{x}-\vect{s})\,p(\vect{s}|\vect{s}_{i})\,\textrm{d}\vect{s}.
\end{align}
The intractable conditional distribution $p(\vect{s}|\vect{s}_{i})$ is approximated as in~\eqref{eq:uninformative}. By performing the change of variable
\[
    \vect{s} = \frac{\vect{s}_i}{\delta_i} + \frac{\sigma_{\tau_i}}{\delta_i} \vect{z},
    \quad \vect{z} \sim \mathcal{N}_{\mathbb{C}}(\vect{0}, \vect{I}),
\]
we can rewrite~\eqref{eq:likelihood_paradiffuse_v1} as
\begin{equation}
    p(\vect{x}|\vect{s}_{i})
    \approx \tilde{p}(\vect{x}|\vect{s}_{i})
    = \E_{\vect{z}\sim\mathcal{N}_{\mathbb{C}}(\vect{0}, \vect{I})}
    \Big[ p_n\!\Big(\vect{x} - \frac{\vect{s}_i}{\delta_i} - \frac{\sigma_{\tau_i}}{\delta_i} \vect{z} \Big) \Big].
\end{equation}
Approximating this expectation with a single Monte Carlo sample yields
\begin{equation}
    \tilde{p}(\vect{x}|\vect{s}_{i})
    \approx
    p_n\!\Big(\vect{x} - \frac{\vect{s}_i}{\delta_i} - \frac{\sigma_{\tau_i}}{\delta_i} \vect{z} \Big),
    \quad \vect{z}\sim\mathcal{N}_{\mathbb{C}}(\vect{0}, \vect{I}).
\end{equation}
The approximate likelihood score is obtained via the chain rule as
\begin{align}
    \nabla_{\vect{s}_i} \log \tilde{p}(\vect{x}|\vect{s}_i) 
    &\approx - \frac{1}{\delta_i}
    \left. \nabla_{\vect{n}_i} \log p_n(\vect{n}_i) 
    \right|_{\vect{n}_i=\vect{x}-\frac{\vect{s}_i}{\delta_i}-\frac{\sigma_{\tau_i}}{\delta_i} \vect{z}} \nonumber \\
    &\approx - \frac{1}{\delta_i}
    \left. \vect{S}_{\psi^*}^{\vect{n}}(\vect{n}_i, \tau_i) 
    \right|_{\vect{n}_i=\vect{x}-\frac{\vect{s}_i}{\delta_i}-\frac{\sigma_{\tau_i}}{\delta_i} \vect{z}} \nonumber \\
    &= -\frac{1}{\delta_i}\,
    \vect{S}_{\psi^*}^{\vect{n}}\!\Big(\vect{x}-\frac{\vect{s}_i}{\delta_i}-\frac{\sigma_{\tau_i}}{\delta_i} \vect{z}, \tau_i\Big).
\end{align}
The clean speech is then sampled iteratively by applying
\begin{align}\label{eq:disc-rev-post-disc_logp_paradiffuse_v1}
    \vect{s}_{i-1}
    &= \vect{s}_{i}
    - \vect{f}_i \Delta{\tau} \nonumber \\
    &\quad
    + g_{\tau_i}^2 \Big[
        -\frac{\lambda_i}{\delta_i}\,
        \vect{S}_{\psi^*}^{\vect{n}}\!\Big(
            \vect{x}-\frac{\vect{s}_i}{\delta_i}
            -\frac{\sigma_{\tau_i}}{\delta_i} \vect{z},
            \tau_i
        \Big)
        + {\vect{S}_{\psi^*}^{\vect{s}}}(\vect{s}_i, \tau_i)
      \Big]\Delta{\tau} \nonumber \\
    &\quad
    + g_{\tau_i}\sqrt{\Delta{\tau}}\,\bs{\zeta}.
\end{align}  
where $\bs{\zeta}$ denotes standard complex Gaussian noise. The overall \texttt{ParaDiffUSE-IN} algorithm is summarized in Alg.~\ref{fig:paradiffuse}.
\subsection{{Explicit noise modeling: }\texttt{ParaDiffUSE-EN}}
In this section, we follow an approach similar to \texttt{DiffUSEEN} by explicitly modeling the noise as a latent variable. We adopt the observation model in~\eqref{eq:obs_model_explicit} and formulate SE as a joint posterior sampling problem, i.e., $(\vect{s}, \vect{n}) \sim p(\vect{s}, \vect{n} | \vect{x})$. To this end, we proceed as in the E-step of \texttt{DiffUSEEN} and use Gibbs sampling to draw from the joint posterior. Specifically, given a noise estimate $\vect{n}^{*}$, we sample the clean speech from
\[
    p(\vect{s}|\vect{x},\vect{n}^{*}) \propto p(\vect{x}| \vect{s}, \vect{n}^{*})\,p(\vect{s}),
\]
and, given a speech estimate $\vect{s}^{*}$, we sample the noise from
\[
    p(\vect{n}|\vect{x},\vect{s}^{*}) \propto p(\vect{x}| \vect{s}^{*}, \vect{n})\,p(\vect{n}).
\]
Within the reverse iterations, we approximate the posterior scores as follows:
\[
\nabla_{\vect{s}_i} \log p_i(\vect{s}_i|\vect{x},\hat{\vect{n}}_{0,i}) = \nabla_{\vect{s}_i} \log p(\vect{x} | \vect{s}_i, \hat{\vect{n}}_{0,i}) + \nabla_{\vect{s}_i} \log p_i(\vect{s}_i),
\]
when sampling $\vect{s}$, and similarly:
\[
\nabla_{\vect{n}_i} \log p_i(\vect{n}_i|\vect{x},\hat{\vect{s}}_{0,i}) = \nabla_{\vect{n}_i} \log p(\vect{x} | \hat{\vect{s}}_{0,i}, \vect{n}_i) + \nabla_{\vect{n}_i} \log p_i(\vect{n}_i),
\]
when sampling $\vect{n}$. Here, $\vect{n}^{*}$ and $\vect{s}^{*}$ are replaced by the Tweedie estimates $\hat{\vect{n}}_{0,i}$ and $\hat{\vect{s}}_{0,i}$, respectively.

To simplify the likelihood terms $p(\vect{x} | \vect{s}_i, \hat{\vect{n}}_{0,i})$ and $p(\vect{x} | \hat{\vect{s}}_{0,i}, \vect{n}_i)$, we employ the uninformative prior assumption. Specifically, we derive:
\begin{align} 
p\left(\vect{x} | \vect{s}_i, \hat{\vect{n}}_{0,i}\right) &= \int p\left(\vect{x}, \vect{s} | \vect{s}_i, \hat{\vect{n}}_{0,i}\right) \,\mathrm{d}\vect{s}, \\
&= \int p(\vect{x} | \vect{s}, \hat{\vect{n}}_{0,i})\, p\left(\vect{s} | \vect{s}_i\right)\,\mathrm{d}\vect{s}.
\end{align}
Using $p(\vect{x}|\vect{s},\hat{\vect{n}}_{0,i}) = \mathcal{N}_{\mathbb{C}}(\vect{s}+\hat{\vect{n}}_{0,i},\sigma_r^2\vect{I})$ together with \eqref{eq:uninformative}, we perform the following {approximation}
\begin{equation}\label{eq:pseudo_ll_speech_paradiffusev2}
p(\vect{x} | \vect{s}_{i}, \hat{\vect{n}}_{0,i}) \approx \tilde{p}(\vect{x} | \vect{s}_{i}, \hat{\vect{n}}_{0,i}) = \mathcal{N}_{\mathbb{C}}\left(\frac{\vect{s}_i}{\delta_{i}} + \hat{\vect{n}}_{0,i}, \Big(\frac{\sigma_{\tau_i}^2}{\delta_{i}^2}  + \sigma_r^2\Big)\vect{I}\right).
\end{equation}
The corresponding likelihood score is given by:
\begin{equation}\label{eq:score_pseudo_ll_speech_paradiffusev2}
\nabla_{\vect{s}_{i}} {\log\,} \tilde{p}\left(\vect{x} | \vect{s}_{i}, \hat{\vect{n}}_{0,i}\right) = \frac{1}{\delta_i}\Big[\frac{\sigma^2_{\tau_i}}{\delta_i^2} + \sigma_r^2 \Big]^{-1}\Big(\vect{x}-\frac{\vect{s}_i}{\delta_i}-\hat{\vect{n}}_{0,i}\Big).
\end{equation}
An analogous derivation applies to $p(\vect{x} | \hat{\vect{s}}_{0,i}, \vect{n}_i)$, yielding:
\begin{equation}\label{eq:pseudo_ll_noise_paradiffusev2}
\tilde{p}(\vect{x} | \hat{\vect{s}}_{0,i}, \vect{n}_i) = \mathcal{N}_{\mathbb{C}}\left(\frac{\vect{n}_i}{\delta_{i}} + \hat{\vect{s}}_{0,i}, \Big(\frac{\sigma_{\tau_i}^2}{\delta_{i}^2} + \sigma_r^2\Big)\vect{I}\right),
\end{equation}
with its score:
\begin{equation}\label{eq:score_pseudo_ll_noise_paradiffusev2}
\nabla_{\vect{n}_{i}} {\log\,} \tilde{p}\left(\vect{x} | \hat{\vect{s}}_{0,i}, \vect{n}_{i}\right) = \frac{1}{\delta_i}\Big[\frac{\sigma^2_{\tau_i}}{\delta_i^2} + \sigma_r^2 \Big]^{-1}\Big(\vect{x}-\frac{\vect{n}_i}{\delta_i}-\hat{\vect{s}}_{0,i}\Big).
\end{equation}

Finally, the clean speech and noise are iteratively sampled using the following updates:
\begin{equation}\label{eq:disc-rev-post-disc_logp_paradiffuse_s}
\begin{split}
    {\vect{s}}_{{i-1}} = {\vect{s}}_{i} - \vect{f}_i\Delta{\tau} + g_{\tau_i}^2 \Big[\lambda_i\nabla_{\vect{s}_i} \log \tilde{p}(\vect{x} | \vect{s}_i, \hat{\vect{n}}_{0,i}) \\ +\, \vect{S}^{\vect{s}}_{\psi^*}(\vect{s}_i, \tau_i)\Big]\Delta{\tau} + g_{\tau_i}\sqrt{\Delta{\tau}}\bs{\zeta},
\end{split}
\end{equation}
and
\begin{equation}\label{eq:disc-rev-post-disc_logp_paradiffuse_n}
\begin{split}
    {\vect{n}}_{{i-1}} = {\vect{n}}_{i} - \vect{f}_i\Delta{\tau} + g_{\tau_i}^2 \Big[\lambda_i\nabla_{\vect{n}_i} \log \tilde{p}(\vect{x} | \vect{n}_i, \hat{\vect{s}}_{0,i}) \\ +\, \vect{S}_{\psi^*}^{\vect{n}}(\vect{n}_i, \tau_i)\Big]\Delta{\tau} + g_{\tau_i}\sqrt{\Delta{\tau}}\bs{\zeta}.
\end{split}
\end{equation}
At the end of the sampling, we apply Wiener filtering as a post-processing step to improve the SE performance and reinforce 
consistency with the observed mixture. Algorithm~\ref{alg:paradiffuse_en} summarizes the full procedure. Note that, although the uninformative prior used here for likelihood approximation is not accurate, it enables faster inference. In contrast, the approach used in concurrent works~\cite{chung2023parallel, 11464305} requires differentiating the score model with respect to the sampling variable to obtain the likelihood gradient~\cite{chung2023diffusion}, which adds a non-negligible computational overhead at inference time. 

{Table~\ref{tab:algos_comparison} summarizes the differences between the diffusion-based algorithms discussed in Sections~\ref{sec:speech_diff_implicit_nmf_noise}, \ref{sec:speech_diff_explicit_nmf_noise}, and~\ref{sec:paradiffuse}.}   

\begin{algorithm}[t!]
\caption{\texttt{ParaDiffUSE-EN}}  \label{alg:paradiffuse_en}
\begin{algorithmic}[1]

\Require $\vect{x}, N, \lambda, \sigma_r^2$
\vspace{0.2cm}
    \State ${\vect{s}}_N \sim \mathcal{N}_{\mathbb{C}}(\vect{x}, \vect{I}), {\vect{n}}_N \sim \mathcal{N}_{\mathbb{C}}(\vect{x}-\vect{s}_N, \vect{I}), \Delta \tau \gets \frac{1}{N}$
    \vspace{0.2cm}
    \For{$i=N,\ldots, 1$}
    
    \vspace{0.2cm}
    \State $\tau_i\gets i \Delta$, $\lambda_i \gets \textsc{scheduler}_\lambda(i )$
        \vspace{0.2cm}

        \For{$\vect{a}\in \{ \vect{s}, \vect{n}$ \}}
        
        \vspace{0.2cm}
        \State \text{Compute} $\{ \bs{\mu}^{\mathsf{back}}(\vect{a}_{i}), \bs{\Sigma}^{\mathsf{back}}_i\}$ using \eqref{eq:mu_sigma_update_prior} 
        with $\vect{h}({\vect{a}}_{i}) = \vect{a}_{i}$
        \vspace{0.2cm}
        \State ${\vect{a}}^{\mathsf{back}}_{i} \gets \bs{\mu}^{\mathsf{back}}(\vect{a}_{i}) + \sqrt{\bs{\Sigma}^{\mathsf{back}}_i}\bs{\zeta}^{a}_p, \quad \bs{\zeta}^{a}_p\sim \normalc$

\vspace{0.2cm}
\State $\hat{\mathbf{a}}_{0,i} \gets \delta_{i}^{-1}\Big({\vect{a}}^{\mathsf{back}}_{i} + \sigma^2_{\tau_i} \vect{S}_{\psi^*}^{\vect{a}}({\vect{a}}^{\mathsf{back}}_{i},\tau_i)\Big)$ 

\EndFor
\vspace{0.2cm}
            
                \State $c_i \gets\dfrac{1}{{\delta}_{i}}\Big[(\dfrac{{\sigma^2_{\tau_i}}}{{{\delta}^2_{i}}} + \sigma_r^2 )\vect{I}\Big]^{-1} $
                \State $\nabla_{{\vect{s}}_i} \log {p} (\vect{x}|{\vect{s}}^{\mathsf{back}}_{i},\hat{\mathbf{n}}_{0,i}) \gets c_i
                             \Big(\vect{x}-({\delta_{i}^{-1}}{{\vect{s}^{\mathsf{back}}_{i}}} + \hat{\mathbf{n}}_{0,i})  \Big) $
                             
    \vspace{0.2cm}
                \State $\nabla_{{\vect{n}}_{\tau}} \log {p} (\vect{x}|\hat{\mathbf{s}}_{0,i},{\vect{n}}^{\mathsf{back}}_{i}) \gets c_i
                             \Big(\vect{x}-({\delta_{i}^{-1}}{{\vect{n}^{\mathsf{back}}_{i}}} + \hat{\mathbf{s}}_{0,i})  \Big) $
\vspace{0.2cm}
                 
                \State $\vect{s}_{i-1} \gets {\vect{s}}^{\mathsf{back}}_{i}  + {\lambda_{i}}{g_{i}^2} {\nabla_{\vect{s}_i} \log \tilde{p} (\vect{x}|\vect{s}^{\mathsf{back}}_i,\hat{\mathbf{n}}_{0,i})} \Delta{\tau}$ 
    \vspace{0.2cm}
                \State $\vect{n}_{i-1} \gets {\vect{n}}^{\mathsf{back}}_{i}  + {\lambda_{i}}{g_{i}^2} {\nabla_{\vect{n}_i} \log \tilde{p} (\vect{x}|\hat{\mathbf{s}}_{0,i},\vect{n}^{\mathsf{back}}_i)} \Delta{\tau}$
                             
    \vspace{0.2cm}            

\vspace{0.2cm}
    \EndFor
\vspace{0.2cm}
\State \textit{Wiener filtering:}
\vspace{0.2cm}
    
    \State ${\vect{a}_{0}} \gets \frac{|\vect{a}_{0}|}{\sqrt{|\vect{s}_{0}|^{\odot 2}+|\vect{n}_{0}|^{\odot 2}}}|\vect{x}|\,\mathrm{e}^{i \angle(\vect{a}_{0})}$,~ $\forall \vect{a}\in \{\vect{s}, \vect{n} \}$ 
\vspace{0.2cm} 
    \State \Return $\hat{\vect{s}}={\vect{s}}_{0}, \hat{\vect{n}}={\vect{n}}_{0}$
\end{algorithmic}
\end{algorithm}

{\begin{table*}[t]
\footnotesize
\centering
\setlength{\tabcolsep}{3pt} 
\begin{tabularx}{\linewidth}{@{}l *{5}{>{\centering\arraybackslash}X}@{}}
\toprule
 & Noise modeling with diffusion 
 & Explicit noise modeling 
 & No likelihood approx. 
 & No $\lambda$ hyperparam. 
 & Tweedie for faster inference \\
\midrule
\texttt{UDiffSE} \cite{nortier2023unsupervised}  & \xmark & \xmark & \xmark & \xmark & \xmark \\
\texttt{UDiffSE+} \cite{ayilo2024diffavse}       & \xmark & \xmark & \xmark & \xmark & \cmark \\
\texttt{DEPSE-IL} \cite{sadeghi2025posterior}    & \xmark & \xmark & \xmark & \cmark & \cmark \\
\texttt{DEPSE-TL} \cite{sadeghi2025posterior}    & \xmark & \xmark & \cmark & \cmark & \cmark \\
\midrule
\texttt{DiffUSEEN} (proposed)                    & \xmark & \cmark & \xmark & \xmark & \cmark \\
\texttt{ParaDiffUSE-IN} (proposed)               & \cmark & \xmark & \xmark & \xmark & \cmark \\
\texttt{ParaDiffUSE-EN} (proposed)               & \cmark & \cmark & \xmark & \xmark & \cmark \\
\bottomrule
\end{tabularx}
\caption{Comparison of diffusion-based frameworks for unsupervised speech enhancement.}
\label{tab:algos_comparison}
\vspace{-1em}
\end{table*}}

\section{Experiments and results}\label{sec:exp}

\subsection{Experimental settings}
\noindent\textbf{Baselines.} The proposed algorithms (\texttt{DiffUSEEN}, \texttt{ParaDiffUSE-IN}, \texttt{ParaDiffUSE-EN}) are compared against the previous unsupervised diffusion-based SE algorithms reviewed in Section~\ref{sec:speech_diff_implicit_nmf_noise}, namely \texttt{UDiffSE}~\cite{nortier2023unsupervised}, \texttt{UDiffSE+}~\cite{ayilo2024diffavse}, \texttt{DEPSE-IL}, and \texttt{DEPSE-TL}~\cite{sadeghi2025posterior}. We also include the recurrent variational autoencoder ({RVAE}) for SE presented in~\cite{leglaive2020recurrent, bie2022unsupervised} as an unsupervised VAE-based generative baseline. In addition, we consider one representative model from each of the other groups of SE methods discussed in Section~\ref{sec:intro}. For the unsupervised non-generative group, we select {RemixIT}~\cite{tzinis2022remixit}, a teacher–student-based model. For the supervised generative group, we use {SGMSE+}~\cite{richter2023speech}. {Conv-TasNet}~\cite{luo2019conv} represents the supervised non-generative group. It should be noted that our objective is robustness under domain mismatch rather than achieving state-of-the-art (SOTA) results, so the models used here are not necessarily SOTA.

\vspace{0.2cm}

\noindent\textbf{Dataset.} 
As in the previous works \cite{bie2022unsupervised, sadeghi2025posterior}, we use the Wall Street Journal corpus (WSJ0) \cite{garofolo1993csr} and the Voice Bank corpus (VB) \cite{veaux2013voice} for the training of the algorithms requiring generative speech prior. When training the diffusion-based noise prior for \texttt{ParaDiffUSE}, the QUT-Noise dataset \cite{dean2015qut} or the DEMAND dataset \cite{thiemann2013diverse} is used. The training noises of QUT are collected in kitchen, street, car, and cafeteria, while for the test, they are from living room, cafeteria, car, and street. Regarding the DEMAND dataset, the training noises are from office meeting room, cafeteria, restaurant, subway station, car, metro, busy traffic, and additional synthetically created speech-shaped and babble noises. The test noises are recorded in living room, office space, bus, cafeteria, and public square. For methods requiring noisy speech in the training process, we use the synthetically created noisy speech of WSJ0-QUT \cite{leglaive2020recurrent} or VB-DMD \cite{botinhao2016investigating} datasets, which are, respectively, created by mixing the training set of WSJ0 (25 hours) with QUT-Noise \cite{dean2015qut} or the VB corpus with the DEMAND dataset \cite{thiemann2013diverse}. The corresponding test set of WSJ0-QUT (1.48 hours) and the VB-DMD test set (0.58 hour) are used to evaluate the proposed algorithms and the baselines in matched and mismatched settings. In both training and test sets of WSJ0-QUT, the SNRs are \{-5, 0, 5\} dB. The training SNRs in VB-DMD are \{0, 5, 10, 15\} dB and those of the test are \{2.5, 7.5, 12.5, 17.5\} dB. In VB-DMD, noise signals are added to the speech segment that are effectively active according to the ITU-T P.56 method \cite{itu1993objective}, while in WSJ0-QUT, active speech segments are detected following the ITU-R BS.1770-4 method \cite{itu-r-bs.1770-4} and SNR is computed using loudness K-weighted relative to full scale (LKFS). This latter weighting leads to lower SNR values, compared to the sum of the squared signal coefficients method \cite{leglaive2020recurrent}.

\vspace{0.2cm}

\noindent\textbf{Evaluation metrics.} 
We evaluate the SE performance using standard instrumental evaluation metrics: scale-invariant signal-to-distortion
ratio (SI-SDR) in dB \cite{le2019sdr}, the extended short-time objective intelligibility (ESTOI) measure \cite{jensen2016algorithm} ([0, 1]), and the perceptual evaluation of speech quality (PESQ) score \cite{rix2001pesq} ([-0.5, 4.5]). Additionally we use {the DNS-MOS P.808\footnote{ \href{https://github.com/microsoft/DNS-Challenge/blob/591184a9fcb2cbdec02520fed81a32bbbf9d73ff/DNSMOS/dnsmos_local.py}{https://github.com/microsoft/DNS-Challenge/blob/DNSMOS/dnsmos\_local.py}}  overall quality score \cite{reddy2021dnsmos}, a non-intrusive objective metric ranging from 1 to 5 \cite{reddy2022dnsmos}.}

\vspace{0.2cm}

\noindent\textbf{Model architecture.} The neural network used for the unsupervised diffusion-based algorithms, as well as for {SGMSE+}, is a lightweight variant of the Noise Conditional Score Network (NCSN++) \cite{richter2023speech}, with 5.2 million parameters. To condition the score model on the label (either speech or noise) in the \texttt{ParaDiffUSE} algorithms, we use the FiLM fusion \cite{perez2018film} to inject both the diffusion timestep and label embeddings into the residual blocks of the network. This yields a single joint model with 5.9 million parameters, instead of two separate models with 5.2 million parameters each. For the {RVAE} baseline \cite{bie2022unsupervised}, we adopt a non-causal architecture based on bidirectional LSTMs. For {RemixIT}, we adopt the setup provided in the UDASE challenge baseline\footnote{\href{https://github.com/UDASE-CHiME2023/baseline}{https://github.com/UDASE-CHiME2023/baseline}} \cite{LEGLAIVE2025101685}. The teacher model is a pretrained {Sudo rm -rf} checkpoint \cite{tzinis2020sudo}, trained in a non-generative supervised fashion on Libri1Mix, and the student shares the same architecture. All neural networks are trained from scratch.

\vspace{0.2cm}

\noindent\textbf{Hyperparameters setting.} All signals are sampled at 16~kHz. For all diffusion-based frameworks, the STFT and SDE hyperparameters follow those of {SGMSE+}~\cite{richter2023speech}. Specifically, we use a Hann window of size 510 with a hop length of 128 for STFT computation, and discretize the reverse SDE into $N=30$ steps. The remaining baselines use their default hyperparameters. When the number of training epochs is not specified in the original code, we train for 200 epochs. For {RemixIT}, the teacher model is updated every 30 epochs \cite{10360251}. The {RVAE} model uses a latent space of dimension 16.

Regarding $\lambda_i$, which balances the prior score and the gradient of the noisy-speech log-likelihood, we use a constant value across diffusion steps for \texttt{UDiffSE}, \texttt{UDiffSE+}, \texttt{DiffUSEEN}, and \texttt{ParaDiffUSE-IN}. Concretely, we set $\lambda_i$ to 1.5 (following~\cite{nortier2023unsupervised}), 1.5 (following~\cite{ayilo2024diffavse}), 1.75, and 1, respectively. {The values for \texttt{DiffUSEEN} and \texttt{ParaDiffUSE-IN} are selected based on a grid search over the validation set.} For \texttt{ParaDiffUSE-EN}, we found that setting $\lambda_i$ proportional to the standard deviation of the perturbation kernel \eqref{eq:ai_a0} associated with the forward SDE yields better enhancement. Specifically, we define $\lambda_{i} = \lambda \times \sigma_{\tau_i}$ with $\lambda = 5.75$. For the manually added Gaussian noise $\mathbf{r}$ in \texttt{DiffUSEEN} and \texttt{ParaDiffUSE-EN}, we set $\sigma_r = 5 \times 10^{-4}$ {after grid-searching}. The NMF rank is set to 4 for all diffusion-based frameworks using NMF, following~\cite{nortier2023unsupervised}.

\vspace{0.2cm}
\subsection{Results}
Table~\ref{tab:wsj_vb_merged} presents the average metrics for the different algorithms under matched and mismatched conditions. We boldface and underline the values that represent statistically significant improvements ($p < 0.05$), either as the best or second-best scores, based on paired  {Welch’s} t-tests for related samples. 


\vspace{0.2cm}
\noindent
\textit{1. Impact of explicit noise modeling in unsupervised methods}

\vspace{0.2cm}
We investigate the effect of modeling noise explicitly as a latent variable at inference time for the unsupervised diffusion-based methods, comparing both diffusion-based, i.e., \texttt{ParaDiffUSE-EN}, and NMF-based noise prior models, i.e., \texttt{DiffUSEEN}.

\vspace{0.2cm}
\noindent
\textbf{Diffusion-based speech and noise priors.} Modeling the noise with diffusion without explicitly representing it as a latent variable, as in \texttt{ParaDiffUSE-IN}, leads to notably poorer performance compared to \texttt{ParaDiffUSE-EN}. We observe that \texttt{ParaDiffUSE-IN} consistently lags behind all other diffusion-based SE frameworks across SI-SDR, PESQ, ESTOI, and DNS-MOS metrics in all configurations, except in the matched condition of VB-DMD. This underperformance may also be partially attributed to the approximations made in the score-likelihood computation. Further investigation is needed to better understand the limitations of \texttt{ParaDiffUSE-IN}.

\vspace{0.2cm}
\noindent
\textbf{Diffusion-based noise prior and NMF-based noise prior.} Comparing \texttt{DiffUSEEN} to the other diffusion-based methods that do not require acoustic noise data during training (in contrast to \texttt{ParaDiffUSE-IN} and \texttt{ParaDiffUSE-EN}), we observe that it achieves the best performance in terms of SI-SDR, PESQ, and ESTOI across all conditions and on both datasets. Its overall enhancement quality, as measured by DNS-MOS, is also among the best in this category. This is noteworthy, as \texttt{DiffUSEEN} attains these results with fewer inference iterations than \texttt{UDiffSE}.

\begin{table*}[t!]
\captionsetup{font=footnotesize}
\caption{Average results on WSJ0\textendash QUT (left block) and VB\textendash DMD (right block) under matched and mismatched training conditions (\textbf{overall best}, \underline{second best}). ``Unsup.'' indicates if the SE framework is unsupervised, ``Gen.'' indicates if the method is generative. The upper left panel indicates that the models are trained and evaluated on WSJ0-QUT (matched WSJ0-QUT), the upper right panel indicates that the models are trained and evaluated on VB-DMD (matched VB-DMD). The lower left panel indicates that the models are trained on VB-DMD and evaluated on WSJ0-QUT (mismatched WSJ0-QUT), and the lower right panel indicates that the models are trained on WSJ0-QUT and evaluated on VB-DMD (mismatched VB-DMD).}

\setlength{\tabcolsep}{6pt}
\centering
\scriptsize
\begin{tabular}{@{}c l c c c c c c c c | c c c c c c@{}}
\toprule
\multicolumn{10}{c|}{\textbf{WSJ0-QUT}} & \multicolumn{6}{c}{\textbf{VB-DMD}} \\
\cmidrule(lr){1-10}\cmidrule(l){11-16}
 & \multicolumn{1}{c}{Method} & Unsup. & Gen. & SI-SDR & SI-SIR & SI-SAR & PESQ & ESTOI & DNS-MOS & SI-SDR & SI-SIR & SI-SAR & PESQ & ESTOI & DNS-MOS \\
\midrule
\multirow{12}{*}{\rotatebox[origin=c]{90}{Matched}} 

& Input & -- & -- & -2.60 & -2.60 & 46.66 & 1.83 & 0.50 & 2.69 & 8.45 & 8.45 & 47.51 & 3.02 & 0.79 & 3.07 \\
\cmidrule(lr){2-16} 
& SGMSE+ \cite{richter2023speech} & \xmark & \cmark & \underline{8.53} & \underline{24.65} & \underline{8.72} & \underline{2.61} & \underline{0.76} & \textbf{3.63} & 17.16 & 28.89 & 17.65 & \textbf{3.51} & \textbf{0.85} & \textbf{3.51} \\
& Conv-TasNet \cite{luo2019conv} & \xmark & \xmark & \textbf{12.63} & \textbf{29.88} & \textbf{12.75} & \textbf{2.86} & \textbf{0.83} & \underline{3.57} & \textbf{19.26} & \textbf{32.36} & 19.80 & \underline{3.40} & \textbf{0.85} & 3.31 \\
& RemixIT \cite{tzinis2022remixit} & \cmark & \xmark & 4.14 & 22.45 & 4.32 & 2.21 & 0.68 & 3.06 & 1.58 & 9.22 & 3.32 & 2.13 & 0.64 & 2.90 \\

\cmidrule(lr){2-16} 

& RVAE \cite{bie2022unsupervised} & \cmark & \cmark & 6.22 & 14.18 & 7.70 & 2.33 & 0.64 & 3.25 & 17.03 & 27.63 & 17.85 & 3.18 & 0.81 & 3.29 \\
& \texttt{UDiffSE} \cite{nortier2023unsupervised} & \cmark & \cmark & 4.35 & 9.63 & 6.24 & 2.20 & 0.61 & 3.06 & 13.62 & 20.08 & 15.43 & 3.34 & \underline{0.82} & 3.31 \\
& \texttt{UDiffSE+} \cite{ayilo2024diffavse} & \cmark & \cmark & 3.24 & 12.14 & 4.23 & 2.21 & 0.58 & 3.19 & 10.40 & 22.45 & 11.04 & 3.30 & 0.77 & 3.27 \\
& \texttt{DEPSE-IL} \cite{sadeghi2025posterior} & \cmark & \cmark & 3.25 & 9.63 & 4.85 & 2.20 & 0.59 & 3.11 & 10.48 & 19.70 & 11.54 & 3.35 & 0.78 & 3.26 \\
& \texttt{DEPSE-TL} \cite{sadeghi2025posterior} & \cmark & \cmark & 3.53  &   5.72 &   8.05  &  2.17  &  0.61  & 3.04  & 13.84 & 16.36 & 18.36 & \underline{3.41} & \underline{0.82} & 3.31  \\
& \texttt{DiffUSEEN} & \cmark & \cmark & 5.37 & 9.69 & 7.97 & 2.33 & 0.67 & 3.16 & 14.32 & 17.40 & 18.56 & \underline{3.43} & \underline{0.83} & 3.33 \\

\cmidrule(lr){2-16} 

& \texttt{ParaDiffUSE-IN} & \cmark & \cmark &  2.02  &   4.86  &   6.42  &  2.16  &  0.56  &  3.24  &  13.55 & 20.72 & 15.38 & \underline{3.41} & 0.80 & 3.33  \\

& \texttt{ParaDiffUSE-EN} & \cmark & \cmark & \underline{8.49}  &   21.54  &    \underline{8.90}  &  \underline{2.61} &  0.73  & 3.44 & \underline{17.85} & \underline{29.65} & 18.41 & \underline{3.40} & \textbf{0.84} & \textbf{3.42} \\

\midrule
\multirow{11}{*}{\rotatebox[origin=c]{90}{Mismatched}} 

& SGMSE+ \cite{richter2023speech} & \xmark & \cmark & 3.02 & 12.79 & 3.86 & 2.08 & 0.61 & \textbf{3.59} & 10.06 & 12.43 & \textbf{17.13} & 3.29 & \underline{0.82} & \underline{3.30} \\
& Conv-TasNet \cite{luo2019conv} & \xmark & \xmark & \textbf{5.83} & \underline{17.07} & \textbf{6.36} & \underline{2.21} & 0.63 & 3.07 & 11.13 & 15.22 & \underline{15.90} & 3.28 & \textbf{0.83} & \underline{3.27} \\
& RemixIT \cite{tzinis2022remixit} & \cmark & \xmark & 3.35 & \textbf{19.88} & 3.62 & 2.13 & \textbf{0.65} & 2.94 & 1.17 & 7.04 & 3.38 & 2.10 & 0.63 & 2.86 \\

\cmidrule(lr){2-16} 

& RVAE \cite{bie2022unsupervised} & \cmark & \cmark & \underline{4.98} & 11.66 & \textbf{6.75} & \underline{2.17} & 0.59 & 3.10 & 12.77 & \textbf{37.64} & 12.85 & 3.11 & 0.76 & 3.28 \\
& \texttt{UDiffSE} \cite{nortier2023unsupervised} & \cmark & \cmark & 1.66 & 8.51 & 3.14 & 2.10 & 0.56 & 3.07 & \underline{14.08} & 21.66 & 15.35 & \underline{3.31} & \underline{0.81} & \underline{3.29} \\
& \texttt{UDiffSE+} \cite{ayilo2024diffavse} & \cmark & \cmark & 0.06 & 8.72 & 1.21 & 2.05 & 0.52 & 3.13 & 10.87 & \underline{28.27} & 11.02 & \underline{3.30} & 0.77 & 3.26 \\
& \texttt{DEPSE-IL} \cite{sadeghi2025posterior} & \cmark & \cmark & 0.24  &  6.63  &    1.91  &  2.06 &  0.53  &  3.05  &  11.14 & 24.32 & 11.53 & \textbf{3.33} & 0.78 & 3.23 \\
& \texttt{DEPSE-TL} \cite{sadeghi2025posterior} & \cmark & \cmark & 2.66  &  5.96  &   \underline{6.07}  &  \underline{2.16}  &  0.59  &  3.05  & \underline{13.85} & 17.15 & \textbf{17.43} & \textbf{3.34} & \underline{0.81} & \underline{3.28}  \\
& \texttt{DiffUSEEN} & \cmark & \cmark & \underline{4.59} & 11.79 & \underline{6.15} & \textbf{2.31} & \underline{0.64} & 3.23 & \textbf{14.61} & 19.06 & \textbf{17.28} & \textbf{3.35} & \underline{0.82} & \underline{3.30} \\

\cmidrule(lr){2-16} 

& \texttt{ParaDiffUSE-IN} & \cmark & \cmark & -0.39 &   3.18  &   2.78  &  1.98  &   0.50  &    3.17  &  8.52 & 10.04 & \underline{15.79} & 3.26 & 0.77 & 3.26 \\

& \texttt{ParaDiffUSE-EN} & \cmark & \cmark & 3.65  & 11.34 &   4.93  &  \underline{2.17}  &   0.60  &  \underline{3.27} & 10.71 & 12.82 & \textbf{17.65} & \textbf{3.34} & \textbf{0.83} & \textbf{3.35} \\

\bottomrule
\end{tabular}
\label{tab:wsj_vb_merged}
\end{table*}

Another clear pattern emerges when comparing \texttt{DiffUSEEN} with \texttt{UDiffSE+}. {Both algorithms use a diffusion model as the speech prior and a Gaussian distribution with an NMF-based covariance to model the noise. The key difference is that, in the proposed \texttt{DiffUSEEN} algorithm, the noise component of the noisy speech is treated as a latent variable, whereas this is not the case in \texttt{UDiffSE+}.} 
\texttt{DiffUSEEN} substantially reduces artifacts, as indicated by the SI-SAR scores (for instance, in the mismatched condition on the WSJ0-QUT test set, we obtain 1.21~dB for \texttt{UDiffSE+} vs.\ 6.15~dB for \texttt{DiffUSEEN}, while in the matched condition on the VB-DMD test set, we obtain 11.02~dB for \texttt{UDiffSE+} vs.\ 17.28~dB for \texttt{DiffUSEEN}). This means that, although \texttt{DiffUSEEN}'s SI-SIR may degrade, the higher SI-SAR maintains a favorable balance with SI-SIR, leading to an overall improvement in SI-SDR. We conjecture that, compared to \texttt{UDiffSE+}, explicitly modeling noise as a latent variable in \texttt{DiffUSEEN} allows us to better capture the interaction between speech and noise. Even though NMF may not perfectly model the acoustic noise in either algorithm, this explicit noise representation helps avoid treating parts of the speech signal as noise (i.e., it removes less energy where speech is present), thereby reducing distortion, as reflected in the improved SI-SAR scores. {We conclude that the proposed methodology in \texttt{DiffUSEEN}, namely EM inference with alternating sampling of speech and noise in the E-step, improves enhancement performance compared to \texttt{UDiffSE+}, where EM inference is performed with only clean-speech sampling in the E-step.}

\vspace{0.2cm}
\noindent
\textbf{Diffusion-based vs. NMF-based noise prior.} In the matched scenario, learning the noise prior with a diffusion model and explicitly modeling the noise as a latent variable, as in \texttt{ParaDiffUSE-EN}, consistently outperforms \texttt{DiffUSEEN}. Both SI-SIR and SI-SAR of \texttt{ParaDiffUSE-EN} are improved compared to \texttt{DiffUSEEN} and the other unsupervised methods, and the same holds for the remaining metrics. For example, in the matched condition on WSJ0-QUT, \texttt{ParaDiffUSE-EN} achieves SI-SIR and SI-SAR scores of 21.54~dB and 8.90~dB, respectively, whereas \texttt{DiffUSEEN} attains 9.69~dB and 7.97~dB, and \texttt{UDiffSE+} 12.14~dB and 4.23~dB. Thus, in the matched scenario, \texttt{ParaDiffUSE-EN} is able to more effectively remove noise while introducing fewer artifacts than the unsupervised baselines, {which is also reflected in its high DNS-MOS score, indicating good overall quality.} This supports the hypothesis that, at least in the matched setting and when noise is explicitly modeled as a latent variable, (pre-trained) diffusion-based noise priors are preferable to (untrained) NMF-based ones.
 
However, in the mismatched case, \texttt{DiffUSEEN} often outperforms \texttt{ParaDiffUSE-EN}. The comparison suggests that shifts in speech and noise distributions at test time more severely affect the model that relies on data-driven priors for both speech and noise. Consequently, \texttt{ParaDiffUSE-EN} exhibits weaker noise-removal performance in the mismatched scenario, as also reflected by its pronounced SI-SIR drop compared to \texttt{DiffUSEEN}. Since \texttt{ParaDiffUSE-EN} and \texttt{DiffUSEEN} mainly differ in how they model noise (the former uses a noise model pre-trained on noise data and kept frozen at inference, while the latter relies on NMF parameters learned during inference), we hypothesize that this inference-time adaptation provides greater robustness to \texttt{DiffUSEEN} under the considered mismatched conditions. Attempts to fine-tune the \texttt{ParaDiffUSE-EN} noise model during inference on each noisy utterance were inconclusive and are therefore not reported here.

\begin{figure}[h!]
    \centering
    \includegraphics[width=1.0\linewidth]{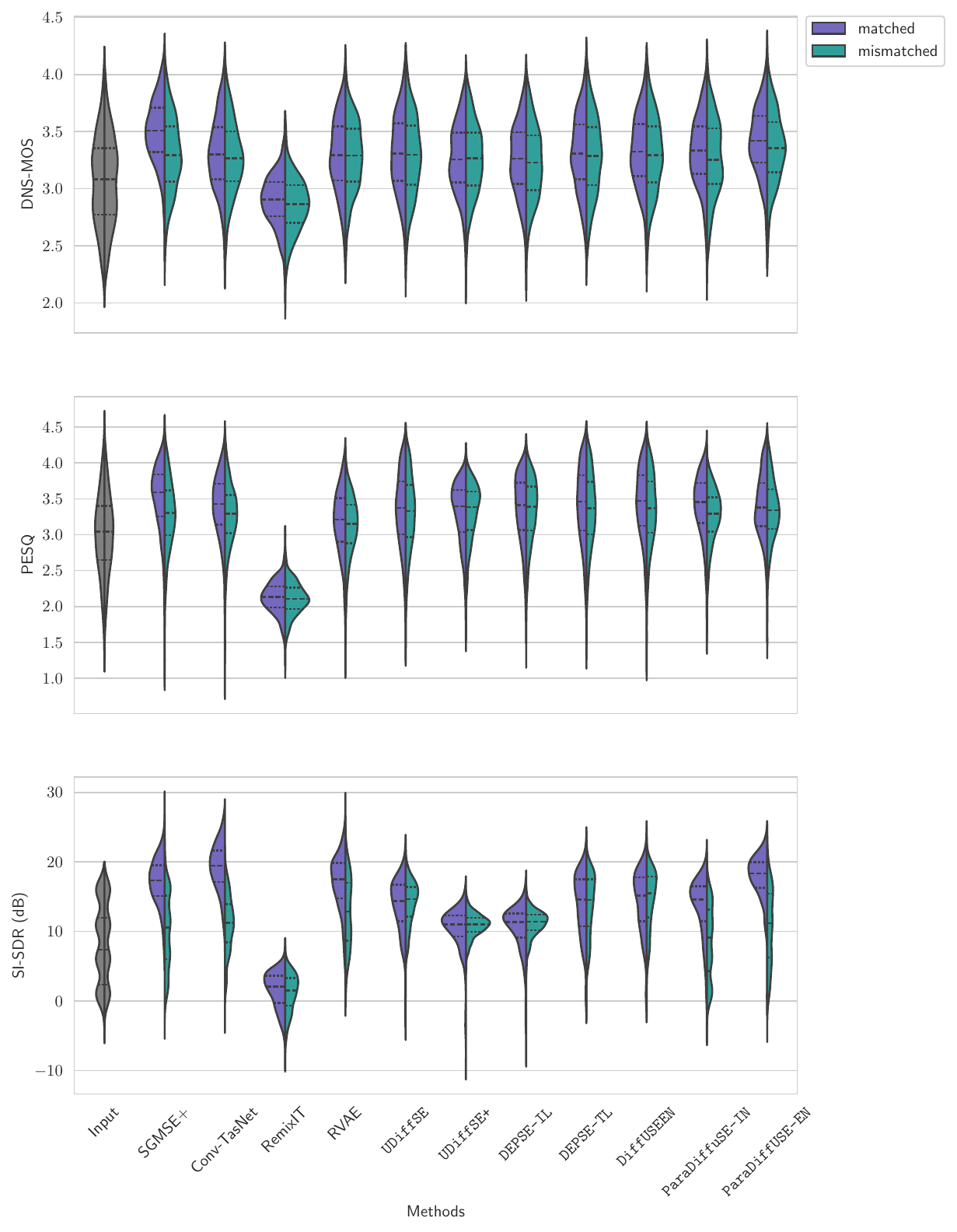}
    \caption{{Violin plots showing the DNS-MOS, PESQ and  SI-SDR distributions for the matched and mismatched conditions on the VB-DMD test set, with dashed and dotted lines indicating the median and quartiles, respectively.}}
    \label{fig:violin_plot}
\end{figure}

\noindent
\textit{2. Comparing matched \& mismatched conditions}

\vspace{0.2cm}
As expected, the supervised methods {SGMSE+} and {Conv-TasNet} achieve the highest performance in matched conditions, except for SI-SDR, SI-SIR, and SI-SAR on VB-DMD, where \texttt{ParaDiffUSE-EN} competes with {SGMSE+} (17.85~dB vs.\ 17.16~dB, 29.65~dB vs.\ 28.89~dB, and 18.41~dB vs.\ 17.65~dB, respectively). {RemixIT} shows low performance on VB-DMD in both matched and mismatched settings. We also note that {RVAE} often outperforms \texttt{DiffUSEEN} in SI-SDR, SI-SIR, and SI-SAR (mainly under matched conditions), whereas \texttt{DiffUSEEN} performs better on quality and intelligibility metrics (PESQ, ESTOI, DNS-MOS). Furthermore, \texttt{DiffUSEEN} remains more robust (less performance degradation) in the mismatched conditions, especially on VB-DMD.

On average, training and testing on different corpora results in a significant performance drop for methods that use noise data during training. Our best baseline in the matched condition, the supervised non-generative method {Conv-TasNet}, is non-negligibly affected by distribution shift, although it remains among the best-performing frameworks for some metrics. For mismatched {WSJ0}-QUT, its SI-SDR drops by 53.84\% (12.63~$\rightarrow$~5.83), and for mismatched VB-DMD by 42.21\% (19.26~$\rightarrow$~11.13). This degradation is also more or less visible on the other metrics. {In particular for PESQ, there is a drop of 22.72\% (2.86 ~$\rightarrow$~ 2.21) for mismatched WSJ0-QUT and 3.52\% (3.40 ~$\rightarrow$~ 3.28) for mismatched VB-DMD}. \texttt{ParaDiffUSE-EN} is similarly affected by such shifts, particularly when training on VB-DMD and evaluating on {WSJ0}-QUT (i.e. mismatched WSJ0-QUT). Specifically, in the mismatched {WSJ0}-QUT scenario, the SI-SDR of \texttt{ParaDiffUSE-EN} drops by 57\% (8.49~$\rightarrow$~3.65), {and the PESQ drops by 20.27\% (2.61~$\rightarrow$~2.17)}, whereas the SI-SDR and PESQ of \texttt{DiffUSEEN} decreases respectively by only 14.52\% (5.37~$\rightarrow$~4.59) and 0.85\%(2.33~$\rightarrow$~2.31). In the mismatched VB-DMD scenario, \texttt{ParaDiffUSE-EN} exhibits a 40\% drop in SI-SDR (17.85~$\rightarrow$~10.71), while \texttt{DiffUSEEN} {even slightly improves, with a 2.13\% increase (14.51~$\rightarrow$~14.82). Regarding PESQ, ESTOI and DNS-MOS metrics, the declines in mismatched VB-DMD scenario are less striking.}

To further highlight the impact of the mismatched data, we show in Fig.~\ref{fig:violin_plot} violin plots of DNS-MOS, PESQ and SI-SDR distributions for different methods on the VB-DMD test set in both matched and mismatched conditions. Similar matched and mismatched distributions for a given method indicate robustness to distribution shift in unseen test conditions and, thus, greater generalization capacity (at least on the datasets used here). {For DNS-MOS, most methods show similar distribution shapes in matched and mismatched scenarios on the VB-DMD test set, suggesting limited perceptual differences according to this non-intrusive metric. However, methods trained with both speech and noise data, such as the \texttt{ParaDiffUSE} variants, RemixIT to a lesser degree, and SGMSE+, show discrepancies between the quartiles of the matched and mismatched distributions. SGMSE+ exhibits larger dispersion in the mismatched scenario, while RemixIT obtains lower DNS-MOS values than the other methods in both scenarios.}

{For the PESQ distributions, as for DNS-MOS, the distribution shape changes only moderately between matched and mismatched scenarios for most algorithms. However, SGMSE+, Conv-TasNet, \texttt{DEPSE-TL}, \texttt{DiffUSEEN}, and \texttt{ParaDiffUSE-IN} show downward shifts in the quartiles under mismatch. This may indicate that PESQ, being intrusive, is more sensitive to mismatch than DNS-MOS, although the perceptual differences on the VB-DMD test set remain limited. A similar plot for the WSJ0-QUT test set, leading to the same conclusion, is reported in Section~D of the supplementary material.}

{In terms of SI-SDR, all algorithms are affected by mismatch, although to different extents. The unsupervised methods \texttt{UDiffSE}, \texttt{UDiffSE+}, \texttt{DEPSE-IL}, \texttt{DEPSE-TL}, and \texttt{DiffUSEEN} appear less affected than \texttt{ParaDiffUSE-EN} and Conv-TasNet. Overall, diffusion-based unsupervised methods that rely on a speech prior, rather than being trained to optimize a reconstruction-fidelity criterion, still show some robustness under mismatch for SI-SDR. This may be due to their independence from noise data during training, together with the explicit noisy-speech consistency enforced through the likelihood term. In contrast, methods trained on noise data may be more sensitive to insufficiently diverse training noise conditions. However, SI-SDR degradation under mismatch does not necessarily imply a strong perceptual difference, and subjective evaluation would be needed to validate this point.}

\begin{table}[t!]
\captionsetup{font=footnotesize}
\caption{Impact of noise-dataset fine-tuning on \texttt{ParaDiffUSE}. Average results on WSJ0-QUT and VB-DMD under mismatched training conditions. For mismatched WSJ0-QUT, the joint diffusion model without fine-tuning is trained on VB-DMD (\texttt{DiffUSEEN} uses only the VB dataset), and fine-tuning is performed on VB-QUT. For mismatched VB-DMD, the joint diffusion model without fine-tuning is trained on WSJ0-QUT (\texttt{DiffUSEEN} uses only the WSJ0 dataset), and fine-tuning is performed on WSJ0-DMD. The highest average within each pair (without fine-tuning vs.\ with fine-tuning) is shown in bold.}

\setlength{\tabcolsep}{6pt}
\centering
\scriptsize
\begin{tabular}{@{}c l >{\centering\arraybackslash}p{.9cm} c c c}
\toprule
\cmidrule(lr){1-6}
 & {Method} & Fine-tuning & SI-SDR & PESQ & ESTOI \\
\midrule
\multirow{6}{.7cm}{\rotatebox[origin=c]{0}{\shortstack{Mismatched\\WSJ0-QUT}}} 
& Input & -- & -2.60 $\pm$ 0.16  & 1.83 $\pm$ 0.02 & 0.50 $\pm$ 0.01 \\
\cmidrule(lr){2-6}
& \texttt{DiffUSEEN} & \xmark & 4.59 $\pm$ 0.22 & {2.31 $\pm$ 0.02} & {0.64 $\pm$ 0.01} \\
\cmidrule(lr){2-6}
& \multirow{2}{*}{\texttt{ParaDiffUSE-IN}} 
  & \xmark & -0.39 $\pm$ 0.27 & 1.98 $\pm$ 0.02 & 0.50 $\pm$ 0.01 \\
& & \cmark & \textbf{1.72 $\pm$ 0.27} & \textbf{2.05 $\pm$ 0.02} & \textbf{0.52 $\pm$ 0.01}\\
\cmidrule(lr){2-6}
& \multirow{2}{*}{\texttt{ParaDiffUSE-EN}} 
  & \xmark & 3.65 $\pm$ 0.29 & 2.17 $\pm$ 0.03 & 0.60 $\pm$ 0.01 \\
& & \cmark & \textbf{5.61 $\pm$ 0.20} & \textbf{2.22 $\pm$ 0.02} & \textbf{0.62 $\pm$ 0.01} \\
\midrule 
\multirow{5}{.7cm}{\rotatebox[origin=c]{0}{\shortstack{Mismatched\\VB-DMD}}} 
& Input & -- & 8.45 $\pm$ 0.20 & 3.02 $\pm$ 0.02 & 0.79 $\pm$ 0.01 \\
\cmidrule(lr){2-6}
& \texttt{DiffUSEEN} & \xmark & 14.61 $\pm$ 0.14 & 3.35 $\pm$ 0.02 & 0.82 $\pm$ 0.00 \\
\cmidrule(lr){2-6}
& \multirow{2}{*}{\texttt{ParaDiffUSE-IN}} 
  & \xmark & 8.52 $\pm$ 0.18 & 3.26 $\pm$ 0.01 & 0.77 $\pm$ 0.00 \\
& & \cmark & \textbf{12.41 ± 0.09} & \textbf{3.35 ± 0.01} & \textbf{0.79 ± 0.00} \\
\cmidrule(lr){2-6}
& \multirow{2}{*}{\texttt{ParaDiffUSE-EN}} 
  & \xmark & 10.71 $\pm$ 0.19 & \textbf{3.34 $\pm$ 0.02} & \textbf{0.83 $\pm$ 0.00} \\
& & \cmark & \textbf{14.96 ± 0.09} & {3.26 ± 0.02} & {0.81 ± 0.00} \\
\bottomrule
\end{tabular}
\label{tab:assess_mismatch_finetune_joint_para}
\end{table}

\vspace{0.2cm}
\noindent
\textit{3. Investigating \texttt{ParaDiffUSE} improvement by fine-tuning}

{We investigate how to improve the semi-supervised \texttt{ParaDiffUSE} variants under mismatched noise conditions, with the goal of performing at least on par with the unsupervised \texttt{DiffUSEEN} method. To this end, we fine-tune the checkpoint of the joint diffusion model trained on WSJ0-QUT using WSJ0-DMD, and the checkpoint trained on VB-DMD using VB-QUT. As shown in Table~\ref{tab:assess_mismatch_finetune_joint_para}, fine-tuning while keeping the speech dataset unchanged and adapting the noise dataset to match the mismatched evaluation test set improves the performance of both \texttt{ParaDiffUSE-EN} and \texttt{ParaDiffUSE-IN} across all metrics compared to the non-fine-tuned models. On mismatched VB-DMD, this improvement allows \texttt{ParaDiffUSE-EN} to perform on par with \texttt{DiffUSEEN} across all metrics. However, on mismatched WSJ0-QUT, although fine-tuning improves the SI-SDR of \texttt{ParaDiffUSE-EN} beyond that of \texttt{DiffUSEEN} (5.61 vs. 4.59), it does not make the PESQ and ESTOI scores of \texttt{ParaDiffUSE-EN} and \texttt{ParaDiffUSE-IN} outperform, or match, those of \texttt{DiffUSEEN}. Future work to improve \texttt{ParaDiffUSE} could focus on diffusion-based test-time adaptation~\cite{barbano2025steerable, chung2024deep}. In addition, Section~B of the supplementary material investigates how speech and noise mismatches individually affect \texttt{ParaDiffUSE-EN} performance, in order to identify which mismatch is more detrimental.}
\begin{table}[ht]
\renewcommand{\arraystretch}{1.5}
\setlength{\tabcolsep}{4pt}
\centering
\scriptsize
\captionsetup{font=footnotesize}
\caption{Average Real Time Factor (RTF) in second (average$\pm$standard error).}
\hfill \break  

\begin{tabular}{lcccc}
\toprule
Methods & {\# Params (M)} & {NFE} & {\# EM} & {RTF ($N=30,\, d=5$) $\downarrow$}\\
\midrule
SGMSE+  \cite{richter2023speech} & {5.21} & {$2\times N$} & {-} & 1.62 $\pm$0.02\\
Conv-TasNet \cite{luo2019conv} & {4.94} & {$1$}  & {-} & 0.02 $\pm$0.01\\
RemixIT \cite{tzinis2022remixit} & {2.95} & {$1$} & {-} & 0.01 $\pm$0.00\\
RVAE \cite{bie2022unsupervised} & {1.02} & {300} & {300} & 37.63 $\pm$0.04\\
\texttt{UDiffSE} \cite{nortier2023unsupervised} & {5.20} & {$2\times N \times d$} & {$d$} & 32.49 $\pm$0.24\\
\texttt{UDiffSE+} \cite{ayilo2024diffavse} & {5.20} & {$2\times N$} & {$N$} & 6.48 $\pm$0.05\\
\texttt{DEPSE-IL} \cite{sadeghi2025posterior}& {5.20} & {$2\times N$} & {$N$} & 6.30 $\pm$0.05\\
\texttt{DEPSE-TL} \cite{sadeghi2025posterior}& {5.20} & {$2\times N$} & {$N$} &  6.31 $\pm$0.05\\
\texttt{DiffUSEEN} & {5.20} & {$2\times N$} & {$N$} & 6.46 $\pm$0.05\\
\texttt{ParaDiffUSE-IN} & {5.94} & {$3\times N$} & {$N$} & 9.82 $\pm$0.07\\
\texttt{ParaDiffUSE-EN} & {5.94} & {$4\times N$} & {$N$} & 13.22 $\pm$0.10\\
\bottomrule
\end{tabular}

\label{tab:rtf}
\end{table}

\vspace{0.2cm}
\noindent
\textit{4. Performance–speed trade-off}

 \begin{figure*}
    \centering
    \includegraphics[width=.8\linewidth]{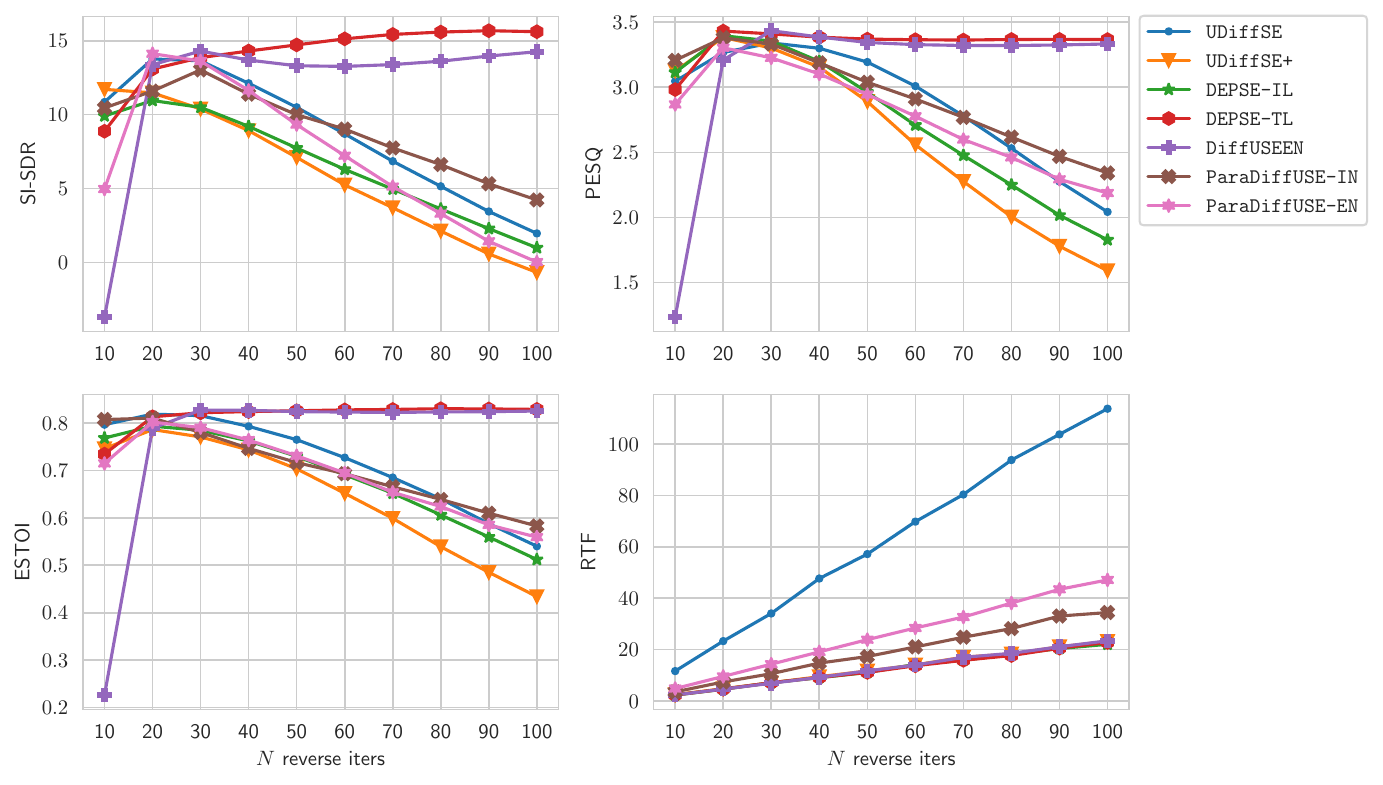}     
    \captionsetup{font=footnotesize}
    \caption{{SE performance on the VB-DMD test set in matched conditions and inference speed as a function of the number of reverse diffusion steps $N$.}}
    \label{fig:perf_speed_tradeoff}
\end{figure*}

\vspace{0.2cm}
{Table~\ref{tab:rtf} reports, for each SE method, the number of neural-network parameters, the number of function evaluations (NFE), the number of EM iterations when applicable, and the average real-time factor (RTF), i.e., the time required to process 1~s of audio. For the unsupervised and semi-supervised diffusion-based algorithms, the RTF is reported with $N=30$ reverse diffusion iterations.}

{The non-iterative methods, Conv-TasNet and RemixIT, are the fastest, processing 1~s of noisy speech in at most 20~ms thanks to a single forward pass through a lightweight network. For diffusion-based methods, including SGMSE+, the Predictor-Corrector sampler requires two score-network evaluations per reverse iteration, one for the Corrector step~\eqref{eq:update_h} and one for the Predictor step~\eqref{eq:mu_sigma_update_prior}. Thus, the NFE of SGMSE+ is $2\times N$. In \texttt{UDiffSE}, each EM E-step runs the full reverse diffusion chain, and Tweedie's formula is not used to shorten inference. Its NFE is therefore $2\times N\times d$, where $d$ is the number of EM iterations. With $N=30$ and $d=5$, as in~\cite{nortier2023unsupervised}, this results in an RTF slightly above half a minute per second of audio.}

{From \texttt{UDiffSE+} onward, inference is accelerated by combining each E-step with a single reverse iteration and Tweedie's formula, followed by the M-step. This leads to $N$ EM iterations and an NFE of $2\times N$. With their respective specificities, \texttt{DEPSE-IL}, \texttt{DEPSE-TL}, and \texttt{DiffUSEEN} follow the same scheme. Consequently, although these methods have a similar model size and number of reverse steps as SGMSE+, they are about four times slower because they perform an M-step after each reverse iteration. Notably, \texttt{DiffUSEEN} has almost the same RTF as \texttt{UDiffSE+} (6.46 vs.\ 6.48), while achieving better SE performance.}

{The \texttt{ParaDiffUSE} methods do not include an M-step to fit the noise component, but instead rely entirely on diffusion-based posterior sampling. In \texttt{ParaDiffUSE-IN}, clean speech is sampled using a Predictor-Corrector step, while the noise score is evaluated only once to compute the likelihood gradient, as shown in line~\ref{lst:line:gradient_ll_noise_paradiffusein} of Algorithm~\ref{alg:ParadiffuSEv1}. This gives an NFE of $3\times N$. In contrast, \texttt{ParaDiffUSE-EN} samples both speech and noise using Predictor-Corrector steps, yielding an NFE of $4\times N$. As a result, \texttt{ParaDiffUSE-EN} is slower than \texttt{ParaDiffUSE-IN} and the other unsupervised diffusion-based methods, except \texttt{UDiffSE}.}

{Finally, RVAE~\cite{bie2022unsupervised} has a relatively high RTF because inference uses a large number of Variational EM iterations, set to 300 in the original configuration. Its NFE depends on the number of epochs used in the Variational E-step with the fine-tuned encoder, here one epoch, and on the length of the STFT sequence processed, here the full noisy STFT spectrogram following the original paper.}

{Figure~\ref{fig:perf_speed_tradeoff} assesses the trade-off between SE performance and inference speed, measured by the average RTF over the test set, for different numbers of reverse diffusion iterations. Most algorithms reach their best performance after 20 or 30 iterations, depending on the metric, after which performance tends to decrease. Although this may seem counter-intuitive, similar behavior has been reported for diffusion-based methods~\cite{yamaguchi2023limitation} and in related iterative problems such as spectrogram inversion~\cite{magron2023spectrogram, wang2019modified}. For $N>30$, the PESQ and ESTOI of \texttt{DiffUSEEN} and \texttt{DEPSE-TL} remain nearly constant, while the SI-SDR of \texttt{DiffUSEEN} first decreases and then increases, and that of \texttt{DEPSE-TL} continues to increase at the cost of higher processing time. Section~C of the supplementary material shows a similar trend under mismatch, with the particularity that, on mismatched WSJ0-QUT, the SI-SDR of \texttt{DiffUSEEN} remains slightly higher than that of \texttt{DEPSE-TL}. As expected, the RTF grows almost linearly with the number of reverse iterations, and methods with larger NFE have higher RTF. The RTFs of \texttt{UDiffSE+}, \texttt{DEPSE-TL}, \texttt{DEPSE-IL}, and \texttt{DiffUSEEN} increase at a similar rate, as they have the same NFE.}

 \begin{figure*}
    \centering
    \includegraphics[width=.9\linewidth]{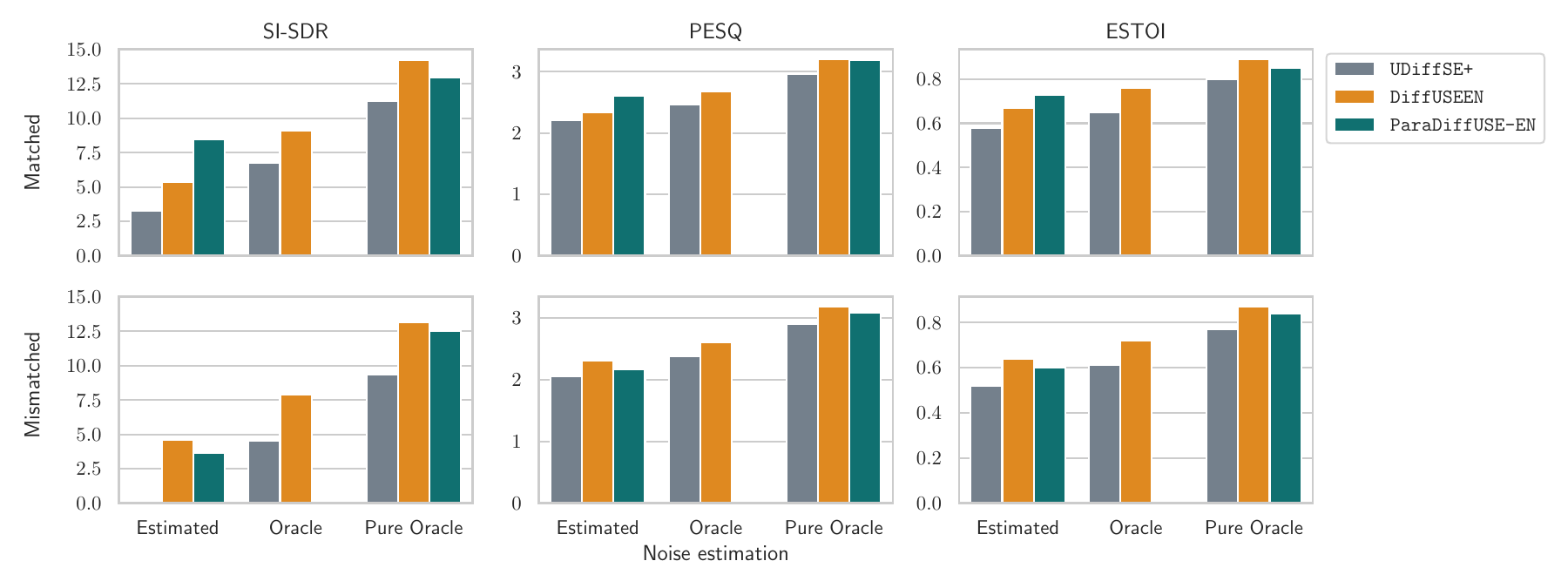}     
    \captionsetup{font=footnotesize}
    \caption{Effect of noise-estimation setting (Estimated / Oracle / Pure Oracle) for \texttt{UDiffSE+}, \texttt{DiffUSEEN}, and \texttt{ParaDiffUSE-EN} on WSJ0–QUT.}
    \label{fig:oracle_plot}
\end{figure*}
\begin{table}[t!]
\renewcommand{\arraystretch}{1.5}
\setlength{\tabcolsep}{4pt}
\centering
\scriptsize
\captionsetup{font=footnotesize}
\caption{Impact of Wiener filtering on \texttt{ParaDiffUSE-EN}. Average results on WSJ0\textendash QUT and VB\textendash DMD under matched and mismatched training conditions. Values are mean $\pm$ standard error. Highest averages are bolded.}
\hfill \break
\begin{tabular}{lcccc} 
\hline 

\textbf{Conditions} & \textbf{Wiener Filter} & SI-SDR  & PESQ & ESTOI \\
\hline\hline

Input (WSJ0-QUT)                  & --         & -2.60 $\pm$ 0.16  & 1.83 $\pm$ 0.02 & 0.50 $\pm$ 0.01 \\
\hline
\multirow{2}{*}{Matched}          & \xmarkred  &  7.46 $\pm$ 0.16  & 2.59 $\pm$ 0.02 & 0.72 $\pm$ 0.01 \\
                                  & \cmarkgreen& \textbf{8.49 $\pm$ 0.19} & \textbf{2.61 $\pm$ 0.02} & \textbf{0.73 $\pm$ 0.01} \\
\hline
\multirow{2}{*}{Mismatched}       & \xmarkred  &  2.86 $\pm$ 1.21  & 2.16 $\pm$ 0.12 & 0.57 $\pm$ 0.04 \\
                                  & \cmarkgreen& \textbf{3.65 $\pm$ 0.29} & \textbf{2.17 $\pm$ 0.03} & \textbf{0.60 $\pm$ 0.01} \\
\hline
\hline

Input (VB-DMD)                    & --         &  {8.45 $\pm$ 0.20} & {3.02 $\pm$ 0.02} & {0.79 $\pm$ 0.01} \\
\hline
\multirow{2}{*}{Matched}          & \xmarkred  & {15.42 $\pm$ 0.09} &	{3.29 $\pm$ 0.01} & {0.79 $\pm$ 0.00} \\
                                  & \cmarkgreen& {\textbf{17.85 $\pm$ 0.11}}	& {\textbf{3.40 $\pm$ 0.02}} &	{\textbf{0.84 $\pm$ 0.00}} \\
\hline
\multirow{2}{*}{Mismatched}       & \xmarkred  & {10.10 $\pm$ 0.17} & {3.24 $\pm$ 0.01} & {0.77 $\pm$ 0.00} \\
                                  & \cmarkgreen& {\textbf{10.71 $\pm$ 0.19}} &	{\textbf{3.34 $\pm$ 0.02}}	& {\textbf{0.83 $\pm$ 0.00}} \\

\hline
\end{tabular}
\label{tab:wsj_vb_wwf}
\end{table}

\vspace{0.2cm}
\noindent
\textit{5. Ablation studies}

\vspace{0.2cm}
\noindent
\textbf{Wiener filter post-processing.}
We perform an ablation study to evaluate the impact of the Wiener filter on the output of the \texttt{ParaDiffUSE-EN} algorithm (no improvement was observed when applying it to \texttt{DiffUSEEN}). The results in Table~\ref{tab:wsj_vb_wwf} show that this post-processing step consistently improves performance across test conditions. On WSJ0-QUT, adding the Wiener filter yields SI-SDR gains of about 1.0~dB in the matched case (7.46~$\rightarrow$~8.49~dB) and 0.8~dB in the mismatched case (2.86~$\rightarrow$~3.65~dB), along with small but systematic increases in PESQ  and ESTOI. 

On VB-DMD, the effect is even more pronounced in the matched setting, where the Wiener filter brings a 2.43~dB SI-SDR improvement (15.42~$\rightarrow$~17.85~dB), a PESQ increase of 0.11 (3.29~$\rightarrow$~3.40), and an ESTOI gain of 0.05 (0.79~$\rightarrow$~0.84). In the mismatched setting, the gains remain consistent. Overall, these results indicate that the Wiener filter systematically refines the \texttt{ParaDiffUSE-EN} estimates, with particularly strong benefits in matched conditions and clear, though more moderate, improvements under distribution shift.

\vspace{0.2cm}
\noindent
\textbf{Estimated vs.\ oracle noise.}
In Figure~\ref{fig:oracle_plot}, we analyze how accurately the methods recover clean speech under different noise-estimation settings on {WSJ0}-QUT. On the x-axis, \textit{Estimated} denotes the realistic case where no ground-truth noise is provided and the algorithm relies on its own noise estimate. \textit{Oracle} (applicable only to NMF-based methods) corresponds to providing the optimal NMF parameters $\vect{W}$ and $\vect{H}$ computed from the ground-truth noise spectrogram.\footnote{We use \texttt{sklearn.decomposition.NMF} with a rank of 4.} \textit{Pure Oracle} means that no noise parameter estimation is performed and the ground-truth noise STFT is directly supplied to the method; this setting represents an upper bound when noise is perfectly known. Note that, in the mismatched condition, the SI-SDR bar for \texttt{UDiffSE+} is barely visible in the \textit{Estimated} case because its value is very small (0.06~dB). As already observed in Table~\ref{tab:wsj_vb_merged}, in the matched case \texttt{ParaDiffUSE-EN} outperforms \texttt{DiffUSEEN}, but this is no longer true under mismatched conditions ({potentially} due to different scale factors, $\lambda_i$). Across test scenarios, as the noise estimate improves, \texttt{DiffUSEEN} tends to produce a better enhanced speech signal and eventually surpasses \texttt{ParaDiffUSE-EN}. One may expect that a more expressive parametric noise model than NMF could further improve results in the \textit{Estimated} setting. 
\begin{table}[t!]
\captionsetup{font=footnotesize}
\caption{Impact of joint versus separate diffusion modeling. Average results on WSJ0\textendash QUT under matched and mismatched training conditions, using either separate diffusion models for speech and noise or a joint (shared) diffusion model. Values are mean $\pm$ standard error. Highest averages within each pair (separate vs.\ joint) are bolded.}

\setlength{\tabcolsep}{6pt}
\centering
\scriptsize
\begin{tabular}{@{}c l c c c c}
\toprule
\cmidrule(lr){1-6}
 & {Method} & Joint & SI-SDR & PESQ & ESTOI \\
\midrule
\multirow{5}{*}{\rotatebox[origin=c]{0}{Matched}} 
& Input & -- & -2.60 $\pm$ 0.16  & 1.83 $\pm$ 0.02 & 0.50 $\pm$ 0.01 \\
\cmidrule(lr){2-6}
& \multirow{2}{*}{\texttt{ParaDiffUSE-IN}} 
  & \xmarkred & \textbf{4.15 $\pm$ 0.21} & \textbf{2.31 $\pm$ 0.02} & \textbf{0.60 $\pm$ 0.01} \\
& & \cmarkgreen & 2.02 $\pm$ 0.24 & 2.16 $\pm$ 0.02 & 0.56 $\pm$ 0.01 \\
\cmidrule(lr){2-6}
& \multirow{2}{*}{\texttt{ParaDiffUSE-EN}} 
  & \xmarkred & \textbf{9.19 $\pm$ 0.16} & \textbf{2.68 $\pm$ 0.02} & \textbf{0.76 $\pm$ 0.01} \\
& & \cmarkgreen & 8.49 $\pm$ 0.19 & 2.61 $\pm$ 0.02 & 0.73 $\pm$ 0.01 \\
\midrule
\multirow{4}{*}{\rotatebox[origin=c]{0}{Mismatched}} 
& \multirow{2}{*}{\texttt{ParaDiffUSE-IN}} 
  & \xmarkred & \textbf{0.41 $\pm$ 0.29} & \textbf{2.09 $\pm$ 0.02} & \textbf{0.52 $\pm$ 0.01} \\
& & \cmarkgreen & -0.39 $\pm$ 0.27 & 1.98 $\pm$ 0.02 & 0.50 $\pm$ 0.01 \\
\cmidrule(lr){2-6}
& \multirow{2}{*}{\texttt{ParaDiffUSE-EN}} 
  & \xmarkred & 3.62 $\pm$ 0.29 & \textbf{2.28 $\pm$ 0.02} & \textbf{0.60 $\pm$ 0.01} \\
& & \cmarkgreen & \textbf{3.65 $\pm$ 0.29} & 2.17 $\pm$ 0.03 & \textbf{0.60 $\pm$ 0.01} \\
\bottomrule
\end{tabular}
\label{tab:assess_joint_sep_para}
\end{table}

\vspace{0.2cm}
\noindent
\textbf{Joint vs. separate noise and speech diffusion models.}
Table~\ref{tab:assess_joint_sep_para} presents an ablation study comparing joint (shared) versus separate speech and noise diffusion models for \texttt{ParaDiffUSE}-based algorithms. On WSJ0-QUT, slightly better scores are obtained when training separate diffusion models, particularly when comparing \texttt{ParaDiffUSE-IN-sep} to \texttt{ParaDiffUSE-IN}, and a similar trend is observed on VB-DMD. However, this comes at the cost of training two networks (2~$\times$~5.2M parameters), whereas a single joint model (5.9M parameters) shares parameters between speech and noise and keeps the model size comparable to our other approaches that do not use a diffusion prior for the noise. Overall, the joint model offers a better trade-off between performance and model complexity.

\section{Conclusion}\label{sec:conc}
We have proposed new diffusion-based frameworks for unsupervised {and semi-supervised} single-channel SE that explicitly model acoustic noise as a latent variable in the noisy mixture, contrary to prior work where noise is implicitly integrated out at inference. This explicit noise modeling allows for a better mixture consistency and more efficient likelihood guidance. More specifically, we introduced \texttt{DiffUSEEN}, which jointly samples speech and noise while keeping an NMF-parameterized Gaussian prior for the noise, and \texttt{ParaDiffUSE-EN}, which learns a conditional diffusion model that serves as a shared prior for both speech and noise. {We also proposed \texttt{ParaDiffUSE-IN}, a variant that, unlike \texttt{ParaDiffUSE-EN}, implicitly accounts for noise during posterior sampling.}

Experiments on WSJ0--QUT and VB-DMD show that explicit noise sampling consistently improves unsupervised diffusion-based SE, regardless of whether the noise prior is NMF-based or diffusion-based. In matched conditions, \texttt{ParaDiffUSE-EN} achieves the best overall quality and intelligibility scores among unsupervised and  and  semi-supervised methods, and is competitive with supervised baselines. Under mismatched conditions, \texttt{DiffUSEEN} is more robust: it exhibits smaller performance drops than its diffusion-based counterpart and than several supervised reference systems, while remaining fully unsupervised with respect to noise data. Ablation studies confirm that Wiener post-filtering systematically refines \texttt{ParaDiffUSE-EN} outputs and that a joint speech–noise diffusion model offers a favorable trade-off between performance and parameter count compared to separate models.

Future work includes designing more expressive, yet efficient, noise priors; reducing the computational cost of posterior sampling to approach real-time operation; {developing effective diffusion-based test-time adaptations for \texttt{ParaDiffUSE} for robustness in mismatched conditions}; and extending the proposed frameworks to multi-channel, audio-visual, and other speech restoration scenarios.


\section{Acknowledgment}
Experiments presented in this paper were carried out using the Grid'5000 testbed, supported by a scientific interest group hosted by Inria and including CNRS, RENATER and several Universities as well as other organizations (see \url{https://www.grid5000.fr}). 

\bibliographystyle{IEEEtran_abbrev}
\bibliography{mybib}

@article{lemercier2025diffusion,
  title={Diffusion models for audio restoration: A review},
  author={Lemercier, Jean-Marie and Richter, Julius and Welker, Simon and Moliner, Eloi and V{\"a}lim{\"a}ki, Vesa and Gerkmann, Timo},
  journal={IEEE Signal Processing Magazine},
  volume={41},
  number={6},
  pages={72--84},
  year={2025},
  publisher={IEEE}
}

@article{richter2023speech,
  title={Speech enhancement and dereverberation with diffusion-based generative models},
  author={Richter, Julius and Welker, Simon and Lemercier, Jean-Marie and Lay, Bunlong and Gerkmann, Timo},
  journal={IEEE/ACM Transactions on Audio, Speech, and Language Processing},
  year={2023},
  publisher={IEEE}
}

@inproceedings{song2021scorebased,
  title={Score-Based Generative Modeling through Stochastic Differential Equations},
  author={Yang Song and Jascha Sohl-Dickstein and Diederik P Kingma and Abhishek Kumar and Stefano Ermon and Ben Poole},
  booktitle={International Conference on Learning Representations (ICLR)},
  year={2021}
}

@inproceedings{song2019generative,
  title={Generative Modeling by Estimating Gradients of the Data Distribution},
  author={Song, Yang and Ermon, Stefano},
  booktitle={Advances in Neural Information Processing Systems},
  pages={11895--11907},
  year={2019}
}

@inproceedings{
  song2022solving,
  title={Solving Inverse Problems in Medical Imaging with Score-Based Generative Models},
  author={Yang Song and Liyue Shen and Lei Xing and Stefano Ermon},
  booktitle={International Conference on Learning Representations},
  year={2022},
  url={https://openreview.net/forum?id=vaRCHVj0uGI}
}

@inproceedings{dean2015qut,
  title={The {QUT-NOISE-SRE} protocol for the evaluation of noisy speaker recognition},
  author={Dean, David and Kanagasundaram, Ahilan and Ghaemmaghami, Houman and Rahman, Md Hafizur and Sridharan, Sridha},
  booktitle={Proceedings of Interspeech},
  pages={3456--3460},
  year={2015},
}

@inproceedings{rix2001pesq,
  title={Perceptual evaluation of speech quality ({PESQ})-a new method for speech quality assessment of telephone networks and codecs},
  author={Rix, Antony W and Beerends, John G and Hollier, Michael P and Hekstra, Andries P},
  booktitle={IEEE international conference on acoustics, speech, and signal processing. Proceedings (ICASSP)},
  volume={2},
  pages={749--752},
  year={2001},
}

@article{fevotte2009nonnegative,
  title={Nonnegative matrix factorization with the Itakura-Saito divergence: With application to music analysis},
  author={F{\'e}votte, C{\'e}dric and Bertin, Nancy and Durrieu, Jean-Louis},
  journal={Neural computation},
  volume={21},
  number={3},
  pages={793--830},
  year={2009},
  publisher={MIT Press}
}

@article{wang2018supervised,
  title={Supervised speech separation based on deep learning: An overview},
  author={Wang, DeLiang and Chen, Jitong},
  journal={IEEE/ACM Transactions on Audio, Speech, and Language Processing},
  volume={26},
  number={10},
  pages={1702--1726},
  year={2018},
  publisher={IEEE}
}

@article{luo2019conv,
  title={{Conv-TasNet}: Surpassing ideal time--frequency magnitude masking for speech separation},
  author={Luo, Yi and Mesgarani, Nima},
  journal={IEEE/ACM transactions on audio, speech, and language processing},
  volume={27},
  number={8},
  pages={1256--1266},
  year={2019},
  publisher={IEEE}
}

@inproceedings{lu2022conditional,
  title={Conditional diffusion probabilistic model for speech enhancement},
  author={Lu, Yen-Ju and Wang, Zhong-Qiu and Watanabe, Shinji and Richard, Alexander and Yu, Cheng and Tsao, Yu},
  booktitle={IEEE International Conference on Acoustics, Speech and Signal Processing (ICASSP)},
  pages={7402--7406},
  year={2022},
}

@article{garofolo1993csr,
  title={{CSR-I (WSJ0)} complete {LDC93S6B}},
  author={Garofolo, John and Graff, David and Paul, Doug and Pallett, David},
  journal={Web Download. Philadelphia: Linguistic Data Consortium},
  volume={83},
  year={1993}
}

@inproceedings{veaux2013voice,
  title={The voice bank corpus: Design, collection and data analysis of a large regional accent speech database},
  author={Veaux, Christophe and Yamagishi, Junichi and King, Simon},
  booktitle={2013 international conference oriental COCOSDA held jointly with 2013 conference on Asian spoken language research and evaluation (O-COCOSDA/CASLRE)},
  pages={1--4},
  year={2013},
  organization={IEEE}
}

@article{lemercier2023storm,
  author={Lemercier, Jean-Marie and Richter, Julius and Welker, Simon and Gerkmann, Timo},
  journal={IEEE/ACM Transactions on Audio, Speech, and Language Processing}, 
  title={{StoRM}: A Diffusion-Based Stochastic Regeneration Model for Speech Enhancement and Dereverberation}, 
  year={2023},
  volume={31},
  number={},
  pages={2724-2737},
  doi={10.1109/TASLP.2023.3294692}}

@article{jensen2016algorithm,
  title={An algorithm for predicting the intelligibility of speech masked by modulated noise maskers},
  author={Jensen, Jesper and Taal, Cees H},
  journal={IEEE/ACM Transactions on Audio, Speech, and Language Processing},
  volume={24},
  number={11},
  pages={2009--2022},
  year={2016},
  publisher={IEEE}
}

@inproceedings{le2019sdr,
  title={{SDR}--half-baked or well done?},
  author={Le Roux, Jonathan and Wisdom, Scott and Erdogan, Hakan and Hershey, John R},
  booktitle={IEEE International Conference on Acoustics, Speech and Signal Processing (ICASSP)},
  year={2019}
}

@inproceedings{reddy2021dnsmos,
  title={DNSMOS: A non-intrusive perceptual objective speech quality metric to evaluate noise suppressors},
  author={Reddy, Chandan KA and Gopal, Vishak and Cutler, Ross},
  booktitle={ICASSP 2021-2021 IEEE International Conference on Acoustics, Speech and Signal Processing (ICASSP)},
  pages={6493--6497},
  year={2021},
  organization={IEEE}
}

@inproceedings{reddy2022dnsmos,
  title={{DNSMOS P. 835: A} non-intrusive perceptual objective speech quality metric to evaluate noise suppressors},
  author={Reddy, Chandan KA and Gopal, Vishak and Cutler, Ross},
  booktitle={IEEE International Conference on Acoustics, Speech and Signal Processing (ICASSP)},
  pages={886--890},
  year={2022}}

@inproceedings{nortier2023unsupervised,
  title={Unsupervised speech enhancement with diffusion-based generative models},
  author={Nortier, Bern{\'e} and Sadeghi, Mostafa and Serizel, Romain},
  booktitle={IEEE International Conference on Acoustics, Speech and Signal Processing (ICASSP)},
  year={2024}}

@article{efron2011tweedie,
  title={Tweedie’s formula and selection bias},
  author={Efron, Bradley},
  journal={Journal of the American Statistical Association},
  volume={106},
  number={496},
  pages={1602--1614},
  year={2011},
  publisher={Taylor \& Francis}
}

@inproceedings{thiemann2013diverse,
  title={The diverse environments multi-channel acoustic noise database ({DEMAND}): A database of multichannel environmental noise recordings},
  author={Thiemann, Joachim and Ito, Nobutaka and Vincent, Emmanuel},
  booktitle={Proceedings of Meetings on Acoustics},
  volume={19},
  number={1},
  year={2013},
  organization={AIP Publishing}
}

@inproceedings{ayilo2024diffavse,
	author={Ayilo, Jean-Eudes and Sadeghi, Mostafa and Serizel, Romain and Alameda-Pineda, Xavier},
  booktitle={IEEE International Conference on Acoustics, Speech and Signal Processing (ICASSP)},
  title={Diffusion-based Unsupervised Audio-visual Speech Enhancement},
  year={2025},
  supp = {https://jeaneudesayilo.github.io/fast_UdiffSE/},
  }

@article{sadeghi2025posterior,
  title={Posterior Transition Modeling for Unsupervised Diffusion-Based Speech Enhancement},
  author={Sadeghi, Mostafa and Ayilo, Jean-Eudes and Serizel, Romain and Alameda-Pineda, Xavier},
  journal={IEEE Signal Processing Letters},
  year={2025},
  publisher={IEEE}
}

@inproceedings{botinhao2016investigating,
  title={Investigating {RNN}-based speech enhancement methods for noise-robust text-to-speech},
  author={Botinhao, Cassia Valentini and Wang, Xin and Takaki, Shinji and Yamagishi, Junichi},
  booktitle={9th ISCA speech synthesis workshop},
  pages={159--165},
  year={2016}
}

@article{daras2024survey,
  title={A survey on diffusion models for inverse problems},
  author={Daras, Giannis and Chung, Hyungjin and Lai, Chieh-Hsin and Mitsufuji, Yuki and Ye, Jong Chul and Milanfar, Peyman and Dimakis, Alexandros G and Delbracio, Mauricio},
  journal={arXiv preprint arXiv:2410.00083},
  year={2024}
}

@inproceedings{
meng2024diffusion,
title={Diffusion Model Based Posterior Sampling for  Noisy Linear Inverse Problems},
author={Xiangming Meng and Yoshiyuki Kabashima},
booktitle={The 16th Asian Conference on Machine Learning (Conference Track)},
year={2024},
url={https://openreview.net/forum?id=A07d0PMRqc}
}

@inproceedings{chung2023parallel,
  title={Parallel diffusion models of operator and image for blind inverse problems},
  author={Chung, Hyungjin and Kim, Jeongsol and Kim, Sehui and Ye, Jong Chul},
  booktitle={Proceedings of the IEEE/CVF Conference on Computer Vision and Pattern Recognition (CVPR)},
  pages={6059--6069},
  year={2023}
}

@book{vincent2018audio,
  title={Audio source separation and speech enhancement},
  author={Vincent, Emmanuel and Virtanen, Tuomas and Gannot, Sharon},
  year={2018},
  publisher={John Wiley \& Sons}
}

@article{vincent2011connection,
  title={A connection between score matching and denoising autoencoders},
  author={Vincent, Pascal},
  journal={Neural computation},
  volume={23},
  number={7},
  pages={1661--1674},
  year={2011},
  publisher={MIT Press}
}

@inproceedings{bando2018statistical,
  title={Statistical speech enhancement based on probabilistic integration of variational autoencoder and non-negative matrix factorization},
  author={Bando, Yoshiaki and Mimura, Masato and Itoyama, Katsutoshi and Yoshii, Kazuyoshi and Kawahara, Tatsuya},
  booktitle={2018 IEEE International Conference on Acoustics, Speech and Signal Processing (ICASSP)},
  pages={716--720},
  year={2018},
  organization={IEEE}
}

@article{bie2022unsupervised,
  title={Unsupervised speech enhancement using dynamical variational autoencoders},
  author={Bie, Xiaoyu and Leglaive, Simon and Alameda-Pineda, Xavier and Girin, Laurent},
  journal={IEEE/ACM Transactions on Audio, Speech, and Language Processing},
  volume={30},
  pages={2993--3007},
  year={2022},
  publisher={IEEE}
}

@inproceedings{leglaive2020recurrent,
  title={A recurrent variational autoencoder for speech enhancement},
  author={Leglaive, Simon and Alameda-Pineda, Xavier and Girin, Laurent and Horaud, Radu},
  booktitle={IEEE International Conference on Acoustics, Speech and Signal Processing (ICASSP)},
  year={2020}
}

@article{itu-r-bs.1770-4,
  author = {{Recommendation ITU-R BS.1770-4}},
  title = {Algorithms to measure audio programme loudness and true-peak audio level},
  journal = {International Telecommunication Union ({ITU})},
  year = {2015},
  url = {https://www.itu.int/rec/R-REC-BS.1770-4-201510-S/en}
}

@article{itu1993objective,
  title={Objective measurement of active speech level},
  author={ITU-T, P},
  journal={ITU-T Recommendation},
  year={1993}
}

@inproceedings{shetu2025gan,
  title={{GAN}-based speech enhancement for low snr using latent feature conditioning},
  author={Shetu, Shrishti Saha and Habets, Emanu{\"e}l AP and Brendel, Andreas},
  booktitle={ICASSP 2025-2025 IEEE International Conference on Acoustics, Speech and Signal Processing (ICASSP)},
  pages={1--5},
  year={2025},
  organization={IEEE}
}

@inproceedings{pascual2017segan,
  title={{SEGAN}: Speech Enhancement Generative Adversarial Network},
  author={Pascual, Santiago and Bonafonte, Antonio and Serr{\`a}, Joan},
  booktitle={Proc. Interspeech 2017},
  pages={3642--3646},
  year={2017}
}

@inproceedings{fu2022metricgan,
  title={{MetricGAN-U}: Unsupervised speech enhancement/dereverberation based only on noisy/reverberated speech},
  author={Fu, Szu-Wei and Yu, Cheng and Hung, Kuo-Hsuan and Ravanelli, Mirco and Tsao, Yu},
  booktitle={IEEE International Conference on Acoustics, Speech and Signal Processing (ICASSP)},
  pages={7412--7416},
  year={2022},
  organization={IEEE}
}

@ARTICLE{10745728,
  author={Kalkhorani, Vahid Ahmadi and Wang, DeLiang},
  journal={IEEE/ACM Transactions on Audio, Speech, and Language Processing}, 
  title={{TF-CrossNet}: Leveraging Global, Cross-Band, Narrow-Band, and Positional Encoding for Single- and Multi-Channel Speaker Separation}, 
  year={2024},
  volume={32},
  number={},
  pages={4999-5009},
  doi={10.1109/TASLP.2024.3492803}}

@article{zhang2025composite,
  title={A Composite Predictive-Generative Approach to Monaural Universal Speech Enhancement},
  author={Zhang, Jie and Yan, Haoyin and Li, Xiaofei},
  journal={IEEE Transactions on Audio, Speech and Language Processing},
  year={2025},
  publisher={IEEE}
}

@inproceedings{wang2023tf,
  title={{TF-GridNet}: Making time-frequency domain models great again for monaural speaker separation},
  author={Wang, Zhong-Qiu and Cornell, Samuele and Choi, Shukjae and Lee, Younglo and Kim, Byeong-Yeol and Watanabe, Shinji},
  booktitle={IEEE international conference on acoustics, speech and signal processing (ICASSP)},
  pages={1--5},
  year={2023},
  organization={IEEE}
}

@article{xiang2020parallel,
  title={A parallel-data-free speech enhancement method using multi-objective learning cycle-consistent generative adversarial network},
  author={Xiang, Yang and Bao, Changchun},
  journal={IEEE/ACM Transactions on Audio, Speech, and Language Processing},
  volume={28},
  pages={1826--1838},
  year={2020},
  publisher={IEEE}
}

@INPROCEEDINGS{10096481,
  author={Fujimura, Takuya and Toda, Tomoki},
  booktitle={IEEE International Conference on Acoustics, Speech and Signal Processing (ICASSP)}, 
  title={Analysis Of Noisy-Target Training For {DNN}-Based Speech Enhancement}, 
  year={2023},
  volume={},
  number={},
  pages={1-5},
  doi={10.1109/ICASSP49357.2023.10096481}}

@article{wisdom2020unsupervised,
  title={Unsupervised sound separation using mixture invariant training},
  author={Wisdom, Scott and Tzinis, Efthymios and Erdogan, Hakan and Weiss, Ron and Wilson, Kevin and Hershey, John},
  journal={Advances in neural information processing systems},
  volume={33},
  pages={3846--3857},
  year={2020}
}

@article{tzinis2022remixit,
  title={{RemixIT}: Continual self-training of speech enhancement models via bootstrapped remixing},
  author={Tzinis, Efthymios and Adi, Yossi and Ithapu, Vamsi K and Xu, Buye and Smaragdis, Paris and Kumar, Anurag},
  journal={IEEE Journal of Selected Topics in Signal Processing},
  volume={16},
  number={6},
  pages={1329--1341},
  year={2022},
  publisher={IEEE}
}

@article{alamdari2021improving,
  title={Improving deep speech denoising by noisy2noisy signal mapping},
  author={Alamdari, Nasim and Azarang, Arian and Kehtarnavaz, Nasser},
  journal={Applied Acoustics},
  volume={172},
  pages={107631},
  year={2021},
  publisher={Elsevier}
}

@article{wu2023self,
  title={Self-supervised speech denoising using only noisy audio signals},
  author={Wu, Jiasong and Li, Qingchun and Yang, Guanyu and Li, Lei and Senhadji, Lotfi and Shu, Huazhong},
  journal={Speech Communication},
  volume={149},
  pages={63--73},
  year={2023},
  publisher={Elsevier}
}

@ARTICLE{10360251,
  author={Zmolikova, Katerina and Pedersen, Michael Syskind and Jensen, Jesper},
  journal={IEEE Open Journal of Signal Processing}, 
  title={Masked Spectrogram Prediction for Unsupervised Domain Adaptation in Speech Enhancement}, 
  year={2024},
  volume={5},
  number={},
  pages={274-283},
  doi={10.1109/OJSP.2023.3343343}}

@inproceedings{
chung2023diffusion,
title={Diffusion Posterior Sampling for General Noisy Inverse Problems},
author={Hyungjin Chung and Jeongsol Kim and Michael Thompson Mccann and Marc Louis Klasky and Jong Chul Ye},
booktitle={The Eleventh International Conference on Learning Representations },
year={2023},
}

@inproceedings{11464305,
  author={Yemini, Yochai and Ben-Ari, Rami and Gannot, Sharon and Fetaya, Ethan},
  booktitle={ICASSP 2026 - 2026 IEEE International Conference on Acoustics, Speech and Signal Processing (ICASSP)}, 
  title={Diffusion-Based Unsupervised Audio-Visual Speech Separation in Noisy Environments with Noise Prior}, 
  year={2026},
  pages={22707-22711},
  doi={10.1109/ICASSP55912.2026.11464305}
 }

@inproceedings{mariani2024multisource,
title={Multi-Source Diffusion Models for Simultaneous Music Generation and Separation},
author={Giorgio Mariani and Irene Tallini and Emilian Postolache and Michele Mancusi and Luca Cosmo and Emanuele Rodol{\`a}},
booktitle={The Twelfth International Conference on Learning Representations},
year={2024},
url={https://openreview.net/forum?id=h922Qhkmx1}
}

@article{LEGLAIVE2025101685,
title = {Objective and subjective evaluation of speech enhancement methods in the UDASE task of the 7th CHiME challenge},
journal = {Computer Speech and Language},
volume = {89},
pages = {101685},
year = {2025},
issn = {0885-2308},
doi = {https://doi.org/10.1016/j.csl.2024.101685},
url = {https://www.sciencedirect.com/science/article/pii/S0885230824000688},
author = {Simon Leglaive and Matthieu Fraticelli and Hend ElGhazaly and Léonie Borne and Mostafa Sadeghi and Scott Wisdom and Manuel Pariente and John R. Hershey and Daniel Pressnitzer and Jon P. Barker},
}

@inproceedings{tzinis2020sudo,
  title={Sudo rm-rf: Efficient networks for universal audio source separation},
  author={Tzinis, Efthymios and Wang, Zhepei and Smaragdis, Paris},
  booktitle={2020 IEEE 30th International Workshop on Machine Learning for Signal Processing (MLSP)},
  pages={1--6},
  year={2020},
  organization={IEEE}
}

@inproceedings{perez2018film,
  title={{FiLM}: Visual reasoning with a general conditioning layer},
  author={Perez, Ethan and Strub, Florian and De Vries, Harm and Dumoulin, Vincent and Courville, Aaron},
  booktitle={Proceedings of the AAAI conference on artificial intelligence},
  volume={32},
  number={1},
  year={2018}
}

@article{barbano2025steerable,
  title={Steerable conditional diffusion for out-of-distribution adaptation in medical image reconstruction},
  author={Barbano, Riccardo and Denker, Alexander and Chung, Hyungjin and Roh, Tae Hoon and Arridge, Simon and Maass, Peter and Jin, Bangti and Ye, Jong Chul},
  journal={IEEE Transactions on Medical Imaging},
  volume={44},
  number={5},
  pages={2093--2104},
  year={2025},
  publisher={IEEE}
}

@inproceedings{chung2024deep,
  title={Deep diffusion image prior for efficient ood adaptation in 3d inverse problems},
  author={Chung, Hyungjin and Ye, Jong Chul},
  booktitle={European Conference on Computer Vision},
  pages={432--455},
  year={2024},
  organization={Springer}
}

@inproceedings{magron2023spectrogram,
  title={Spectrogram inversion for audio source separation via consistency, mixing, and magnitude constraints},
  author={Magron, Paul and Virtanen, Tuomas},
  booktitle={2023 31st European Signal Processing Conference (EUSIPCO)},
  pages={36--40},
  year={2023},
  organization={IEEE}
}

@inproceedings{wang2019modified,
  title={A modified algorithm for multiple input spectrogram inversion},
  author={Wang, Dongxiao and Kameoka, Hirokazu and Shinoda, Koichi},
  booktitle={Proc. Interspeech 2019},
  pages={4569--4573},
  year={2019}
}

@inproceedings{
yamaguchi2023limitation,
title={On the Limitation of Diffusion Models for Synthesizing Training Datasets},
author={Shin'ya Yamaguchi and Takuma Fukuda},
booktitle={SyntheticData4ML 2023},
year={2023},
url={https://openreview.net/forum?id=HVrDgZa5hh}
}


\end{document}